\documentclass[usegraphicx,usenatbib]{mn2e}
\usepackage{color}
\usepackage{amssymb}
\usepackage{epsf,multirow}
\usepackage{graphicx}
\usepackage{txfonts}

\def\nar{New Astronomy Review}

\def\aap{A\&A}
\def\apj{ApJ}
\def\apjs{ApJS}
\def\apjl{ApJL}

\def\mnras{MNRAS}
\def\aj{AJ}
\def\nat{Nature}
\def\aaps{A\&A Supp.}

\def\araa{ARA\&A}   
\def\aj{AJ}
\def\araa{ARA\&A}
\def\apj{ApJ}
\def\apjl{ApJ}
\def\apjs{ApJS}

\def\aap{A\&A}

\def\aaps{A\&AS}

\def\mnras{MNRAS}

\def\pasp{PASP}
\def\pasj{PASJ}

\def\aj{AJ}
\def\araa{ARA\&A}
\def\apj{ApJ}
\def\apjl{ApJ}
\def\apjs{ApJS}

\def\aap{A\&A}

\def\aaps{A\&AS}

\def\mnras{MNRAS}

\def\pasp{PASP}
\def\pasj{PASJ}

\def\fcp{Fund. Cosmic Phys.}

\def\kms{~km~s$^{-1}$}
\def\micron{~$\mu$m}

\def\deg{$^{\circ}$}

\def\Ls{~L$_{\sun}$}

\def\arcsec{$^{\prime\prime}$}
\def\n{NGC~}
\def\S{SINFONI}

\def\Ms{~M$_{\sun}$}
\def\Mbh{M$_{\rm BH}$}
\def\Mc{M$_{\rm cold}$}
\def\Mw{M$_{\rm warm}$}

\def\H2{H$_2$}
\def\Ha{H$\alpha$}
\def\HeI{He\,{\sc i}}
\def\HI{H\,{\sc i}}

\def\OIII{[O\,{\sc iii}]}
\def\Pab{Pa$\beta$}
\def\Brg{Br$\gamma$}

\def\FeII{[Fe\,{\sc ii}]}

\title[Molecular gas in the centre of galaxies -- I]{Molecular gas in the centre of nearby galaxies from VLT/SINFONI integral field spectroscopy -- I. Morphology and mass inventory\thanks{Based on observations at the European Southern Observatory (ESO) Very
Large Telescope [083.B-0126(A) and 083.B-0126(B)].}}

\author[Mazzalay et al.]{X. Mazzalay,$^{1}$\thanks{E-mail: ximena@mpe.mpg.de} R. P. Saglia,$^{1,2}$ Peter Erwin,$^{1}$   M. H. Fabricius,$^{1,2}$  S. P. Rusli,$^{1,2}$ \newauthor J. Thomas,$^{1,2}$  R. Bender,$^{1,2}$ M. Opitsch,$^{1}$ N. Nowak,$^{3}$ Michael J. Williams$^1$\\
$^{1}$Max-Planck-Institut f\"ur extraterrestrische Physik, Postfach 1312, 85741 Garching, Germany\\
$^2$Universit\"atssternwarte, Scheinerstrasse 1, 81679 M\"unchen, Germany\\
$^3$Max-Planck-Institut f\"ur Physik, F\"ohringer Ring 6, 80805 M\"unchen, Germany \\
}

\begin{document}

\date{}


\maketitle


\begin{abstract}

We present the first results of an analysis of the properties of the molecular gas in the nuclear regions (${\rm r} \lesssim 300$~pc) of a sample of six nearby galaxies, based on new high spatial resolution observations obtained in the $K$-band with the near-infrared integral field spectrograph \S\ at the Very Large Telescope. We derive two-dimensional distributions of the warm molecular and ionized gas from the \H2, \Brg\ and \HeI\ emission lines present in the spectra of the galaxies. We find a range of morphologies, including bar- and ring-like distributions and either centrally peaked or off-centre emission. The morphologies of the molecular and the ionized gas are not necessarily coincident. The observed emission-line ratios point towards thermal processes as the principal mechanism responsible for the \H2\ excitation in the nuclear and circumnuclear regions of the galaxies, independently of the presence of an active nucleus. We find that a rescaling of the \H2~2.12\micron\ emission-line luminosity by a factor
$\beta \simeq 1200$ gives a good estimate (within a factor of 2) of the total (cold) molecular gas mass. The galaxies of the sample contain large quantities of molecular gas in their centres, with total masses in the $\sim 10^5-10^8$\Ms\ range. Never the less, these masses correspond to less than 3 per cent of the stellar masses derived for the galaxies in these regions, indicating that the presence of gas should not affect black hole mass estimates based on the dynamical modelling of the stars. The high-spatial resolution provided by the \S\ data allowed us to resolve a circumnuclear ring (with a radius of $\sim 270$~pc) in the galaxy \n4536. The measured values of the \Brg\ equivalent width and the \HeI/\Brg\ emission-line ratio suggests that bursts of star formation occurred throughout this ring as recently as 6.5~Myr ago.

\end{abstract}

\begin{keywords}
galaxies: nuclei -- infrared: galaxies -- galaxies: ISM -- ISM: molecules -- line: formation
\end{keywords}


\section{Introduction}

It is widely accepted that supermassive black holes (SMBHs), with masses between $10^6-10^{10}$\Ms, sit at the centre of elliptical galaxies and bulges of spiral galaxies. The observational correlation between the mass of the central black hole (\Mbh) and the bulge mass or the luminosity and the velocity dispersion ($\sigma$) of the galaxy's bulge, strongly suggests a connection between the formation of the central black holes and the formation and evolution of their hosts \citep[e.g.,][]{Kormendy1995, Magorrian1998, Ferrarese2000, Gebhardt2000, Tremaine2002, Marconi2003, Haring2004, Gultekin2009}.
However, the universality of these relationships has not yet been firmly established. This is mainly due to the fact that they are still based on a limited sample of galaxies, biased towards early types.

In this spirit, an observational spectroscopic survey of nearby galaxies, using the near-infrared (NIR) integral field spectrograph \S\ \citep{Eisenhauer2003, Bonnet2004} at the Very Large Telescope (VLT), was undertaken. The survey is aimed at increasing the number of \Mbh\ estimates to explore the \Mbh--bulge relations, especially at the poorly known low- and high-mass ends, and testing the BH-bulge formation scenarios. This paper is part of a series reporting the results of the \S\ program (Nowak et al. 2007, 2008, 2010; Rusli et al. 2011, 2012a, 2012b; Bender et al. in preparation; Erwin et al. in preparation; Saglia et al. in preparation).
A thorough analysis and derivation of black-hole masses via extensive stellar dynamical modelling can be found in these papers.

Of the total sample of galaxies observed with \S, about one fourth exhibit signatures of gas emission in their nuclei. 
The gaseous material is considered to be the primary fuel source of the nuclear activity. It is not only necessary for the formation and growth of the nuclear BHs but it is also a fundamental ingredient for the nuclear and circumnuclear starburst activity of galaxies.
How is the large amount of gas required to fuel the nuclear activity transported to the inner regions of the galaxies? What is the interplay between star formation and central SMBHs?
These are among the important questions that need to be answered in order to improve our understanding of galactic nuclei, and are currently the focus of intensive research.

The fuel necessary to drive the nuclear activity has to be transported from galactic scales ($\sim 10$~kpc) to scales of less than 1~kpc and into the nucleus. Depending on the scales involved, a number of mechanisms that remove angular momentum from the gas, thus bringing it closer to the centre, have been proposed \citep[see review by][and references therein]{Jogee2006}. Large-scale bars (and the star-forming rings sometimes associated with them), as well as tidal interaction and mergers, seem to play a key role in the transport of the gas into the inner kpc. On smaller scales, nuclear nested bars \citep[e.g.,][]{Shlosman1989} and nuclear spirals \citep[e.g.,][]{Englmaier2000, Maciejewski2004a, Maciejewski2004b} are some of the means of removing angular momentum that have been proposed.

The study of the distribution and properties of the molecular gas and star-forming regions in galactic nuclei can provide clues on the mechanisms channelling the gas to the centre and thus fuelling nuclear activity, and can also help establish the link between the starburst and nuclear SMBHs. The \S\ data set is ideal for this task, since it simultaneously provides high-spatial resolution and the advantages of NIR wavelengths, like low dust extinction and a number of emission lines that can be used as diagnostic for the properties of the molecular gas and the nuclear activity.

This is the first of two papers presenting new integral field spectroscopic (IFS) data for 6 of the galaxies that show molecular emission in our \S\ sample of 33 galaxies\footnote{Two further galaxies, \n1316 and \n3368, show molecular emission and were discussed by \citet{Nowak2008, Nowak2010}.}. Both are aimed at characterizing the properties of the nuclear gas and determining its effect on (and possible use in) the estimation of the mass of the central BH. Here we focus on the emission line gas distribution and mass content of the galaxies, considering the implications for the estimation of BH masses. In a second paper (Mazzalay et al. 2012, in preparation) the kinematics of the gas is analysed and compared with that of the stars. The layout of this paper is as follows. In Section~\ref{s_obs} we describe the new \S\ observations and the data reduction process. In Section~\ref{s_morpho} we describe the main properties of the galaxies in the sample and analyse the morphology of the nuclear regions. Section~\ref{s_pconditions} presents the results on the physical conditions of the molecular and ionized gas. An inventory of the total (gas plus star) mass content is given in Section~\ref{s_mass}. A summary and our main conclusions can be found in Section~\ref{s_summary}.

\section{Observations and analysis of the data}\label{s_obs}

\subsection{IFS observations}

The data presented here were gathered as part of our observational campaign to measure SMBH masses using the NIR integral-field spectrograph SINFONI at the 8~m Very Large Telescope UT4. The galaxies were observed in several runs between 2007 and 2009. A short description of the observations can be found in Table~\ref{t_obs}. We covered the 1.95--2.45\micron\ wavelength range with the $K$-band grating using two different spatial samplings: a high-spatial resolution scale of $0.05 \times 0.10~{\rm arcsec}~{\rm pixel}^{-1}$ (hereafter HR data) and a low-resolution scale of $0.125 \times 0.250~{\rm arcsec}^2~{\rm pixel}^{-1}$ (hereafter LR data). The associated fields of view (FOVs) are approximately $3'' \times 3''$ and $8'' \times 8''$, respectively.
The observations were performed following a standard object-sky-object strategy, with individual exposures of 10~min  dithered by a few spatial pixels (or `spaxels'). This procedure allows a good correction of bad pixels and cosmic ray removal. The total on-source exposure times are listed in column~4 of Table~\ref{t_obs}. The high-spatial resolution observations were AO-assisted, employing either the laser guide star (LGS) PARSEC \citep{Rabien2004} or the natural guide star (NGS) mode using the galaxy nucleus, while the low-resolution observations were taken in natural seeing.
In order to assess the AO performance and characterize the point spread function (PSF) due to atmospheric turbulence, we regularly observed a PSF star immediately after the science exposure sequence, in addition to the normal (non-AO) observation of a telluric standard star. The PSF stars for individual galaxy observations were chosen to have optical ($B - R$) colours and $R$-band magnitudes similar to that of the inner $r < 3''$ of the galaxy nuclei, in order to achieve a similar AO correction. Column~6 of Table~\ref{t_obs} lists the FWHM of the PSF star associated with each AO-assisted observation and, for the cases of the non-AO observations, the seeing values at 2.2\micron\ estimated from those given by the Differential Image Motion Monitor (DIMM). Note that, since the DIMM seeing is measured at 0.5\micron, we assumed that the resolution $\theta$ depends on the wavelength as $\theta(\lambda) \sim \lambda^{-1/5}$ in order to have an estimation of the seeing at 2.2\micron. 
Both high- and low-spatial resolution configurations give a spectral resolution ${\rm R}\sim 4000$ at 2.12\micron, corresponding to an instrumental dispersion of $\sigma_{\rm ins} \sim 30$\kms.

\begin{table*}
\caption{Log of observations. The columns show the field of view (FOV) covered by the observations, observing date, on-source exposure time, AO mode (where NGS stands for `natural guide star' and LGS for `laser guide star'), estimated FWHM of the point spread function and position angle of the observation.}
\label{t_obs}
\begin{tabular}{lcccccc}
\hline
Galaxy & FOV & Obs. date & T$_{\rm exp}$ [min] & AO mode & PSF & Obs. P.A.\\
\hline
\n3351 & $3'' \times 3''$ & 2009-04-19 & 40 & LGS & 0.20\arcsec & 10 \\
       & $3'' \times 3''$ & 2009-04-20 & 40 & LGS & 0.18\arcsec & 10 \\
       & $8'' \times 8''$ & 2009-04-21 & 40 & no AO & 0.45\arcsec & 10 \\
\n3627 & $3'' \times 3''$ & 2007-03-21 & 80 & LGS & 0.24\arcsec & $-12$ \\
       & $3'' \times 3''$ & 2007-03-25 & 50 & LGS & 0.27\arcsec & $-12$ \\
       & $8'' \times 8''$ (N) & 2007-03-25 & 80 & no AO & 0.67\arcsec & $-12$ \\
       & $8'' \times 8''$ (S)& 2009-04-22 & 20 & no AO& 0.67\arcsec & $-12$ \\
       & $8'' \times 8''$ (S)& 2009-05-18 & 20 & no AO& 0.59\arcsec & $-12$ \\
\n4501 & $3'' \times 3''$ & 2008-03-12 & 80 & NGS & 0.13\arcsec & 0 \\
\n4536 & $3'' \times 3''$ & 2009-04-19 & 60 & LGS & 0.18\arcsec & 125 \\
       & $3'' \times 3''$ & 2009-04-22 & 80 & LGS & 0.18\arcsec & 125 \\
       & $8'' \times 8''$ & 2009-04-21 & 80 & no AO & 0.74\arcsec & 125 \\
\n4569 & $3'' \times 3''$ & 2008-03-08 & 80 & NGS & 0.15\arcsec & 25 \\
\n4579 & $3'' \times 3''$ & 2008-03-08 & 40 & NGS & 0.15\arcsec& 96 \\
       & $3'' \times 3''$ & 2008-03-09 & 80 & NGS & 0.15\arcsec& 96 \\
\hline
\end{tabular}

\medskip
\end{table*}

\subsection{Data reduction}

The data were reduced using a custom pipeline incorporating ESOREX \citep{Modigliani2007} and SPRED \citep{Schreiber2004, Abuter2006}. The reduction process comprised all standard reduction steps, including bias subtraction, flat-fielding, bad pixel removal, detector distortion and wavelength calibrations, sky subtraction \citep{Davies2007a}, reconstruction of the object data cubes and, finally, telluric and flux calibrations. Through the reduction tasks, the data cubes were re-sampled to a spatial scale of $0.05 \times 0.05~{\rm arcsec}^2~{\rm pixel}^{-1}$ and $0.125 \times 0.125~{\rm arcsec}^2~{\rm pixel}^{-1}$ for the HR and LR data, respectively. 
Telluric correction and flux calibration of the reconstructed data cubes were performed using the telluric standard stars observed at the beginning or the end of the corresponding observing block at an airmass similar to that of the galaxy. Each star spectrum was extracted from an aperture with a diameter of 5--7 times the mean FWHM and its continuum was modelled with a blackbody curve of the same effective temperature and $K$ magnitude as the star.
To test our flux calibration, we used 2MASS images in combination with high resolution HST (NIC2 and NIC3) and William Herschel Telescope (WHT) NIR images available for \n3351, \n4569 and \n4579.
The images were obtained from the HST archive, NASA/IPAC Infrared Science archive and \cite{Knapen2003}.
We find an agreement between the different data sets consistent to 20 per cent.

\subsection{Construction of the 2D maps}

The \S\ data presented here allow us to study the morphology and physical conditions of the circumnuclear molecular gas of the galaxies in the sample. To improve the signal-to-noise ratio (SNR) of the two-dimensional data and, at the same time, preserve the highest possible spatial resolution, we used the adaptive 2D-binning method of \citet{Cappellari2003}. This method makes use of Voronoi tessellations to bin pixels together by accreting new pixels into the bin until a desired SNR is reached. Here the SNR was defined as the ratio between the flux of the particular emission line analysed and the rms of the continuum, optimally weighted as described in section~2.1 of \citet{Cappellari2003}. The chosen SNR threshold depended on each individual galaxy and was selected so as not to compromise spatial resolution in the high SNR regions and, at the same time, be able to extract information from the regions which showed signs of emission but too low SNR to obtain otherwise meaningful line parameters.

In order to extract information about the molecular gas we fitted a Gaussian component to the \H2~2.12\micron\ emission line in the continuum-subtracted spectra of the galaxies, obtaining the central wavelength, width, and intensity of the line at each spatial bin.
This line was selected since it is one of the strongest \H2\ lines observed in the $K$-band range and it is located in a region not contaminated by other spectral features. We used the measured intensity to construct  emission-line flux distribution maps of the \H2\ emission line gas.
The stellar continuum was fitted and subtracted applying the penalized pixel fitting method (pPXF) of \citet{Cappellari2004}. For this we used a stellar template constructed as a positive linear combination of the spectra of six late-type stars observed with the same configuration as the galaxy data (in this way we ensure that the stellar template and the objects spectra have the same instrumental broadening). Particular care was taken in masking regions contaminated by emission lines and/or spurious features (i.e., bad pixels or bad sky-lines subtraction).

Visual inspection of the emission-line profiles showed no significant deviation from a single-Gaussian component for all the galaxies except \n4579. The \H2\ emission-line profiles of this galaxy display strong asymmetries and double components in some regions. In consequence, to obtain a good estimation of the total integrated line-flux for this galaxy, we described its line profiles as the sum of two Gaussian components.
The resulting maps of \H2~2.12\micron\ flux distribution are shown and analysed in the next section.

Additionally, \Brg\ emission was observed in the spectra of \n3351, \n3627 and \n4536, and the \HeI~2.06~\micron\ emission line was present in the spectra of \n4536. Following the same procedure as in the case of \H2, we constructed the flux distribution maps for these lines and included their analysis in the next section.

\begin{figure*}
\centering
\includegraphics[width=0.7\textwidth]{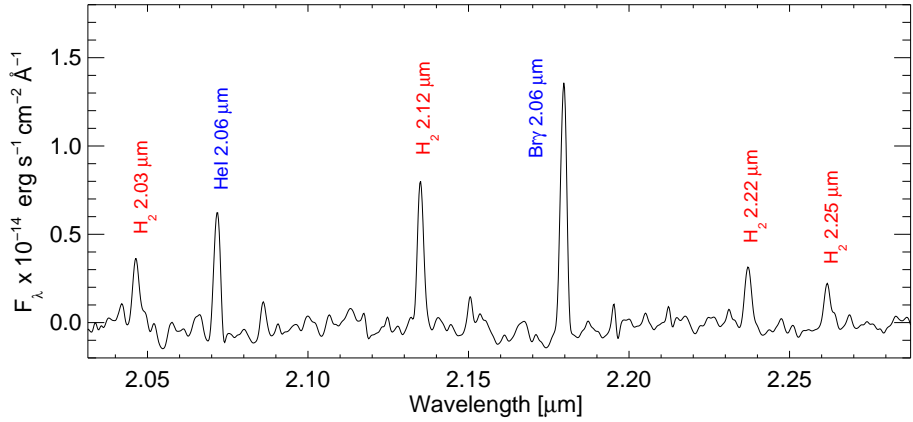}
\caption {Continuum-subtracted \S\ spectrum of the inner $8'' \times 8''$ of \n4536. The labels mark the position of the emission lines identified.}
\label{f_spectrum}
\end{figure*}

As an example, Fig.~\ref{f_spectrum} shows the integrated continuum-subtracted spectrum of the inner $\sim 8''\times 8''$ of \n4536 obtained in the LR configuration. 

\section{Morphology of the circumnuclear gas}\label{s_morpho}

\begin{table*}
\begin{minipage}{155mm}
\caption{Properties of the galaxies of the sample.}
\label{t_prop}
\begin{tabular}{lcccccc}
\hline
				& \n3351    & \n3627    & \n4501    & \n4536    & \n4569    & \n4579\\
\hline
Morphology $^a$			&  SB(r)b   &  SAB(s)b  &  SA(rs)b  & SAB(rs)bc & SAB(rs)ab & SAB(rs)b\\
Nuclear activity	&  --  	    & LINER/Sy2 $^a$      & Sy2 $^a$  & LLAGN $^b$& LLAGN $^c$& LINER/Sy1.9 $^c$\\
Distance (Mpc)			&  10.0  $^d$ & 10.1 $^d$ & 16.5 $^e$ & 14.9 $^d$ & 16.5  $^e$ & 16.5 $^e$ \\
Linear scale (pc arcsec$^{-1}$)	&  49	    &  49     	&  80       &  72       &   80      & 80   \\
Inclination (deg)		&  46       &  65 $^c$  &   60      &  69       &  69        & 40  \\
PA (deg)			&  10       & 178 $^f$  &  140      &  125      & 25         & 95    \\
PA of the stellar bar (deg)	&  112 $^g$ & 161 $^h$ &   --  & -- &   15 $^i$ & 58 $^j$ \\
\hline
\end{tabular}

\medskip
$^{(a)}$Taken from NED. $^{(b)}$\cite{McAlpine2011}. $^{(c)}$\cite{Ho1997}. $^{(d)}$\cite{Freedman2001}. $^{(e)}$Mean Virgo cluster distance \citep{Mei2007}. $^{(f)}$\cite{Casasola2011}. $^{(g)}$\cite{Erwin2005}.  $^{(h)}$\cite{Sheth2002}. $^{(i)}$\cite{Jogee2005}. $^{(j)}$\cite{Garc'ia-Burillo2005}.
\end{minipage}
\end{table*}

All the galaxies discussed in this paper display \H2\ emission lines in their spectra and, in some cases, \Brg\ and \HeI~2.06\micron\ emission were also observed. In this section we describe and analyse the emission-line flux distribution maps of each individual galaxy, as well as relevant previous observations which will allow us to understand the context in which these data reside.
All the \S\ maps are centred at the $K$-band continuum maximum; the continuum was estimated by integrating the signal in the 2.1--2.3\micron\ wavelength range. 
The main properties of the galaxies are summarised in Table~\ref{t_prop}.
The inclination and position angles of the galaxies were derived from a combination of analysis of the shape of outer galaxy isophotes and (where available) large-scale gas kinematics in the literature; details for individual galaxies will be presented in Erwin et al. (in preparation).

\subsection{\n3351}\label{s_3351}

\begin{figure*}
\includegraphics[width=0.7\textwidth]{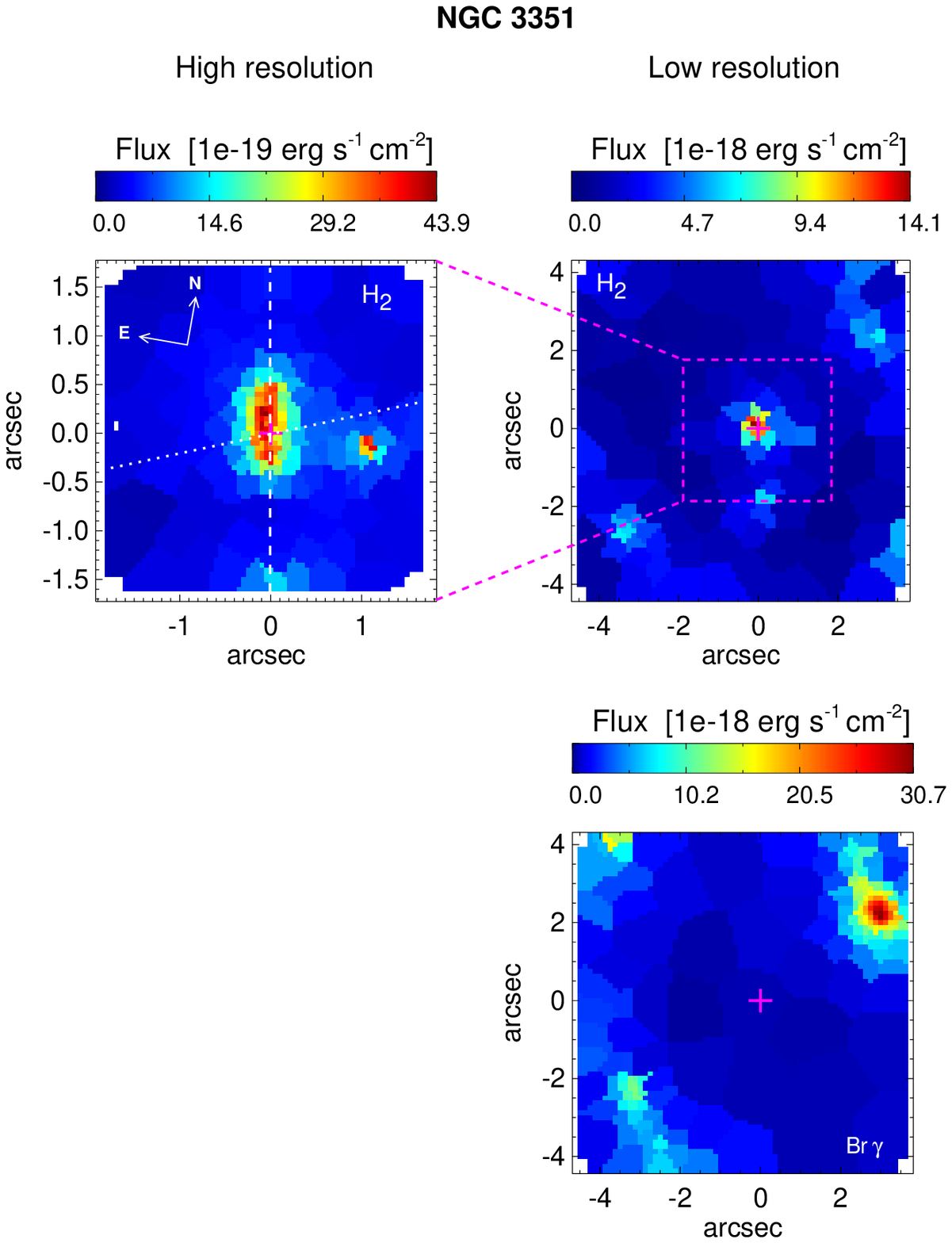}
\caption {Flux distribution of \H2~2.12\micron\ (upper panels) and \Brg\ (lower panel) emission lines observed in \n3351. The panel on the left corresponds to the high-resolution data (${\rm FOV}\sim 3'' \times 3''$) and the panels on the right correspond to the low-resolution data (${\rm FOV}\sim 8'' \times 8''$).
The spatial orientation, indicated in the upper-left panel, is the same for all the panels. The position of the major axis of the galaxy (dashed line) and the orientation of the stellar bar (dotted line) are also indicated.}
\label{f_n3351_maps}
\end{figure*}

\n3351 (M95) is one of the main members of the Leo I group of galaxies and it is classified as an early type barred spiral galaxy, with no signs of activity in its nucleus. It hosts a star-forming nuclear ring with a major axis of $\sim$15\arcsec, which dominates the appearance of the inner region of this galaxy, and a larger ring of H{\sc ii} regions that encircles its stellar bar \citep[e.g.,][]{Rubin1975, Devereux1992, Colina1997, Knapen2002}. Aligned almost perpendicular to the stellar bar, a nuclear `molecular gas bar' $\sim 19''$ long was first reported by \cite{Devereux1992}. Higher spatial resolution observations resolved this into two peaks of CO  emission at a radius of $\sim 7''$, coincident with the circumnuclear ring observed in \Ha\ images \citep{Jogee2005}.

\n3351 was observed with \S\ in two different configurations, with high- and low-resolution spatial samplings, covering  $\sim3'' \times 3''$ and $\sim8'' \times 8''$ FOVs. The \H2\ and \Brg\ flux distribution maps derived from these data sets are presented in Figure~\ref{f_n3351_maps}. 
The HR data shows that the nuclear emission is resolved into two different clumps embedded in diffuse emission (upper-left panel of Fig.~\ref{f_n3351_maps}). The innermost emission is elongated along the major axis of the galaxy, with a relatively constant flux over the inner $\sim 1''$ and a sharp drop to values below 20 per cent of the flux maximum further out. A smaller off-centre source is located at $\sim 1''$ towards the NW of the nucleus. 
These two clumps are not individually resolved in the LR data (upper-right panel of Fig.~\ref{f_n3351_maps}), where we can see an almost symmetric central source connected towards the SW by a low-luminosity ridge to the inner edge of the nuclear ring observed in \Ha\ and UV images \citep{Colina1997, Planesas1997} of this galaxy. Although only part of the ring falls within the \S\ field, the observed \H2\ and \Brg\ distribution found here agrees very well with previous observations of \Ha\ emission-line maps, especially at the location of the hot spots. The brightest \Brg\ emission can be associated with the \Ha\ hot spot labelled as R2 by \cite{Planesas1997} and the emission observed at the east edge of the \S\ field can be associated with the R5, R6 and R7 \Ha\ hot spots \citep[see figure 8 of][]{Planesas1997}. However, the \Brg\ map does not show the nuclear emission observed in both the \H2\ maps presented here and previous \Ha\ images. This is probably due to the intrinsic weakness of the \Brg\ line, which falls below the sensitivity limit of our data. Hints of nuclear \Brg\ emission is observed in the inner $\sim 1''$ of the HR spectra, although it is not strong enough to allow the construction of an emission-line map.

\subsection{\n3627}\label{s_n3627}

\begin{figure*}
\centering
\includegraphics[width=0.7\textwidth]{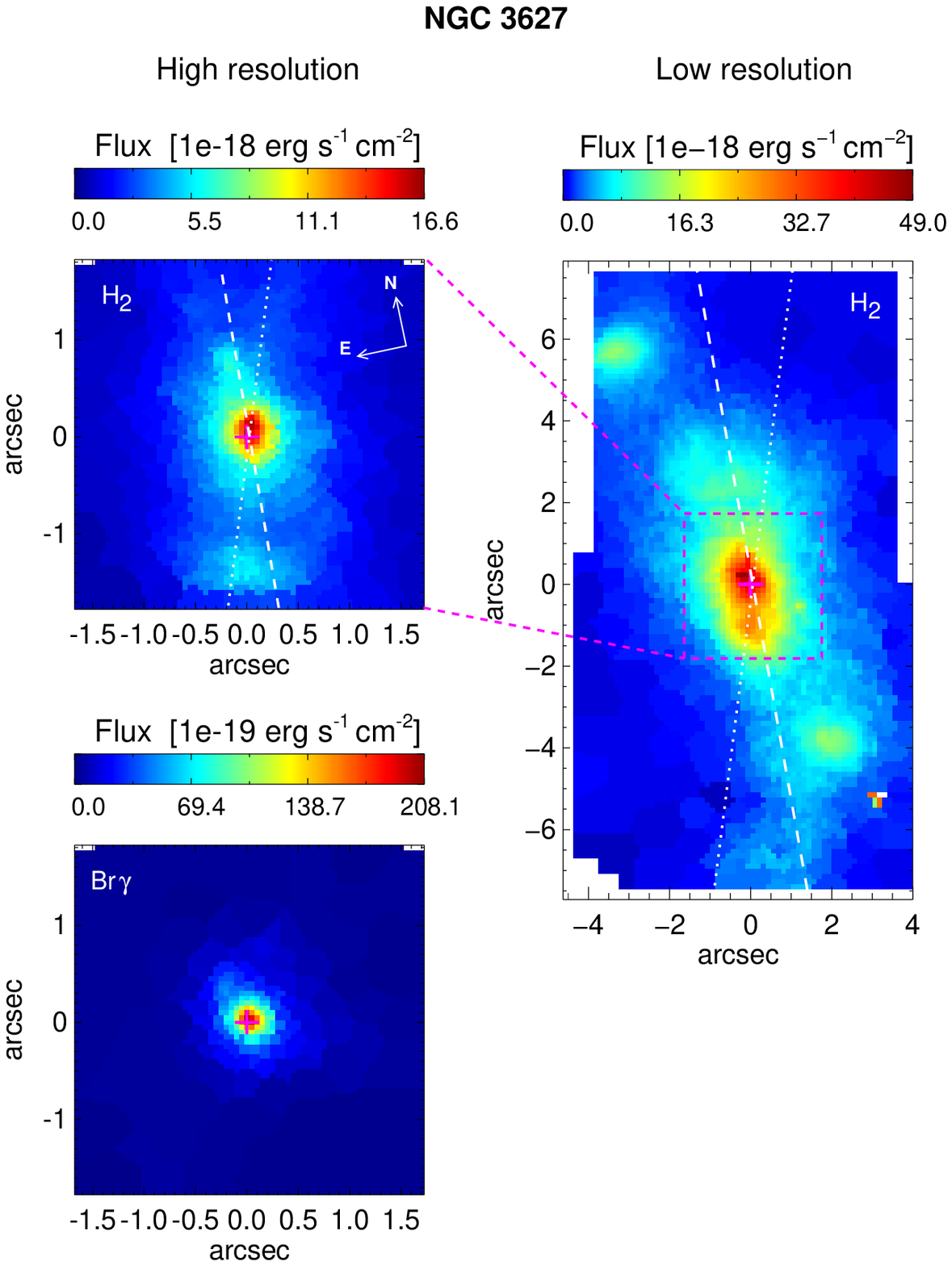}
\caption {Flux distribution maps of the \H2\ and \Brg\ emission lines  observed in \n3627. The panels on the left correspond to the high-resolution data and the panel on the right corresponds to the low-resolution data. The spatial orientation, indicated in the upper-left panel, is the same for all the panels. The dashed line in the \H2\ maps indicates the position of the major axis of the galaxy and the dotted line corresponds to the stellar bar position angle.}
\label{f_n3627_maps}
\end{figure*}

\n3627 (M66) is a high-inclination strongly barred galaxy. It belongs to the Leo Triplet group of galaxies, showing some signatures of past interaction with its neighbour galaxy \n3628. Optical emission-line ratios are inconclusive with respect to the type of nucleus this galaxy harbours. Whether it is a LINER/H II composite or Seyfert 2 like nucleus \citep{Ho1997}, the presence of a variable, flat-spectrum nuclear radio source is consistent with a low-luminosity AGN \citep[LLAGN,][]{Filho2000}. 
Broad-band optical images of \n3627 reveal a pronounced spiral pattern and a central region dominated by several straight dust lanes that run parallel to the host galaxy semi-major axis \citep{Martini2003a}.
The CO emission shows a nuclear peak and extends along the leading edges of the stellar bar, forming two broad peaks at the bar ends, from where the spiral arms trail off from \citep[e.g.,][]{Sheth2002}.

Molecular and recombination lines of hydrogen were observed in the \S\ spectra of \n3627. The left panels of Fig.~\ref{f_n3627_maps} show the flux distribution maps of the \H2~2.12\micron\ and \Brg\ lines in the inner $\sim 3'' \times 3''$ (HR maps) of this galaxy. Additionally, two sets of low-resolution observations were taken with a spatial offset along the $-12$\deg\ position angle, yielding a total FOV of $\sim 8'' \times 15''$. The final \H2\ distribution map obtained from these data (LR map) is presented in the right panel of Fig.~\ref{f_n3627_maps}. No evidence of \Brg\ emission was found at distances larger than a few arcseconds from the nucleus, therefore we only show its distribution derived from the HR data.

As it can be seen from Fig.~\ref{f_n3627_maps}, while the emission of \Brg\ appears strictly confined to the nucleus, within a region of $\lesssim 50$~pc in radius, the morphology of the \H2\ emission in the inner $3'' \times 3''$ is extended along N--S direction. 
Further out, the gas is mainly located along a bar-like structure, with two hot-spots located approximately at distances of 6.5\arcsec\ and 4.5\arcsec\ from the nucleus along a position angle of $\sim 17$\deg\ (from north through east). Additionally, weak emission extending southward is observed beyond the southern hot-spot.
The morphology displayed by the \H2\ emission line is remarkably consistent with the $^{12}$CO(2--1) morphology reported by \cite{Casasola2011} for the inner $20'' \times 20''$ of \n3627 (see their figure~4). 
The position angle of the molecular bar-like structure observed in the innermost regions of \n3627 differs from the position angle of the stellar bar (${\rm PA} = 161$\deg, indicated with a dotted line in Fig.~\ref{f_n3627_maps}) and the large-scale molecular bar \citep[$\rm{PA} = 164$\deg,][]{Sheth2002}. This is because we are looking well within the transition region, where the molecular gas goes from leading the northern part of the stellar bar to lead the southern one \citep[see, for example, figure~1 of][]{Regan2001}.

\subsection{\n4501}\label{s_n4501}

\begin{figure}
\centering
\includegraphics[width=0.75\columnwidth]{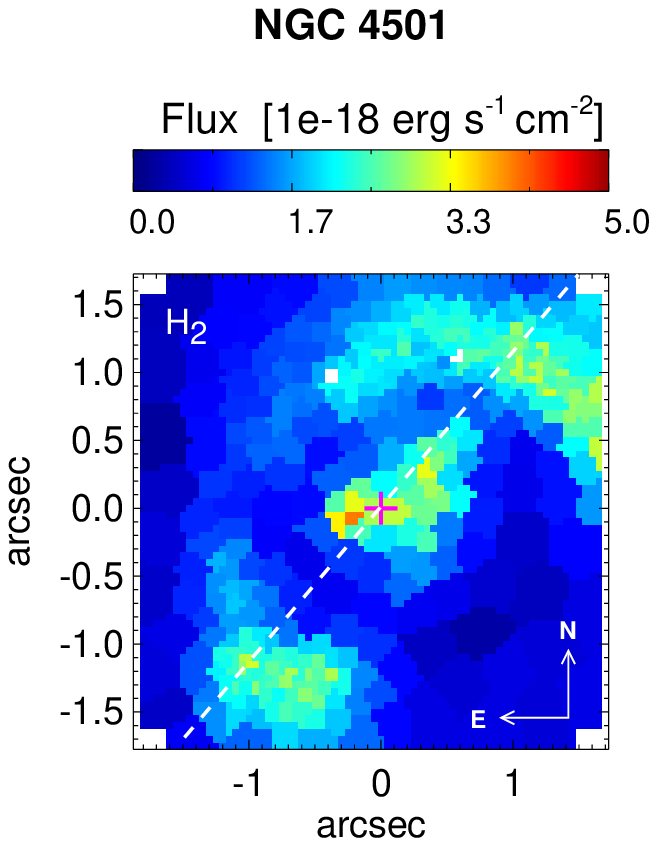}
\caption {Flux distribution map of the \H2\ emission line observed in \n4501. The dashed line represents the orientation of the major axis of the galaxy.}
\label{f_n4501_maps}
\end{figure}

\begin{figure}
\includegraphics[width=\columnwidth]{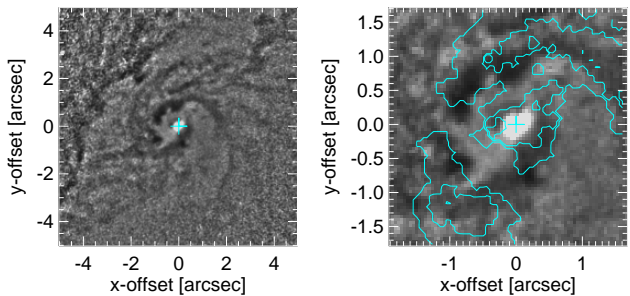}
\caption{Left: Unsharp mask of the HST/WFPC2 F547M image of the inner $10'' \times 10''$ of \n4501. Dust lanes trace spiral patterns around and to the nucleus. Right: Zoom into the inner $3'' \times 3''$ with the \H2\ contours overlaid.}
\label{f_n4501_dust}
\end{figure}

The Seyfert~2 galaxy \n4501 (M88) is one of the largest spiral galaxies in the Virgo cluster region.
$B$-band images presented by \cite{Elmegreen1987} showed multiple external arms and two symmetric inner arms. In the $K'$-band, the spiral structure is continuous within ${\rm r} \lesssim 3$~kpc \citep{Elmegreen1999} and \cite{Carollo1998} were able to observe spiral-like dust lanes (thought to indicate the location of shocks) down to the nucleus, using WFPC2 F606W images.
High-resolution interferometric observations of the $^{12}$CO(1--0) emission in the central 5~kpc of this galaxy were analysed in detail by \cite{Onodera2004}. The CO maps showed spiral arms, associated with the dust lanes found by HST, and a strong central condensation of radius $\sim 5''$ resolved into double peaks separated by 4.7\arcsec, which are located on the `root of the dust spirals'.
Even though \n4501 does not show evidence for a bar, the gas concentration in its centre is relatively high, and more typical of barred than unbarred galaxies \citep{Sakamoto1999b}. 

Figure~\ref{f_n4501_maps} shows the flux distribution map of the \H2~2.12\micron\ emission-line in the  inner $3'' \times 3''$ of \n4501 observed with \S. Two distinct structures can be resolved in the integrated flux map: an asymmetric nuclear component (note that the maximum of the \H2\ emission is offset from the continuum maximum indicated by the plus symbol) surrounded by two arcs that resemble an incomplete ring-like structure. 
The arcs traced by \H2\ towards the NW and SE of the nucleus seems to be located on the inside edges of the two peaks seen in CO by \cite{Onodera2004}, indicating a gradient in the temperature of the molecular gas in the sense that the warm molecular gas (traced by the \H2~2.12\micron\ emission) is located closer to the nucleus than the cold molecular gas (traced by the CO emission). 
A comparison between the \H2\ emitting-gas distribution and the nuclear spiral-like dust lanes exhibit by \n4501 can be seen in Fig~\ref{f_n4501_dust}. This figure shows an unsharp mask of the HST/WFPC2 F547M image (retrieved from the HST archive) of the inner regions of the galaxy, with the \H2\ contours overlaid.
As in the case of the CO emission, a close spatial coincidence between the \H2\ molecular gas and the dust lanes is observed, especially in the north region where both the molecular and dust structures are better defined, suggesting a close relation between the two media down to distances of less than 1\arcsec\ from the nucleus.

\subsection{\n4536}\label{s_n4536}

\n4536 is a large spiral galaxy located in the Virgo cluster. While it has been classified as a SAB(rs)bc galaxy \cite[e.g.,][]{deVaucouleurs1991, Buta2010}, the evidence for the presence of a bar is not yet clear (see for example Pompea \& Rieke 1990; Shaw et al. 1995, Erwin et al. in preparation). 
Although optical line ratios place this galaxy in the H\,{\sc ii} region range \citep{Ho1997}, the presence of high-ionization lines \citep{Satyapal2008} and its X-ray spectrum \citep{McAlpine2011} suggest that this galaxy hosts a low-luminosity AGN in its nucleus. 
Additionally, there is abundant evidence pointing to vigorous star formation taking place in the inner $\sim 20'' \times 30''$ region of this galaxy \citep[e.g.,][]{Puxley1988, Pogge1989b, Telesco1993, Davies1997, Jogee2005}.
This is also a typical `single-peak' type galaxy, with the CO molecular gas being concentrated in a nuclear disc of $\sim 10''$ radius and an unresolved compact core at the nucleus \citep[e.g.,][]{Sofue2003, Jogee2005}. Previous low-resolution (1.6\arcsec) images of \H2~2.12\micron\ emission in the inner 10\arcsec\ of this galaxy were analysed by \cite{Davies1997}. They found a strong NIR continuum peak in the nucleus, but the \H2\ emission has peaks distributed in what seems to be an edge-on ring structure around the nucleus. Their analysis suggests that while active star formation is taking place in the circumnuclear ring, the star formation in its nucleus has come to an end \citep[see also][]{Jogee2005}. Very recently, \citet{Rosenberg2012} presented non-AO $H$-, $J$- and $K$-band \S\ data of the inner $8'' \times 8''$ of \n4536. These authors confirmed the presence of a circumnuclear ring in \n4536, traced mainly by \Brg\ and \FeII~1.26\micron\ emission, while the \H2~2.12\micron\ distribution is more centrally peaked.

\begin{figure*}
\centering
\includegraphics[width=0.7\textwidth,trim= 0 0 0 0, clip=true]{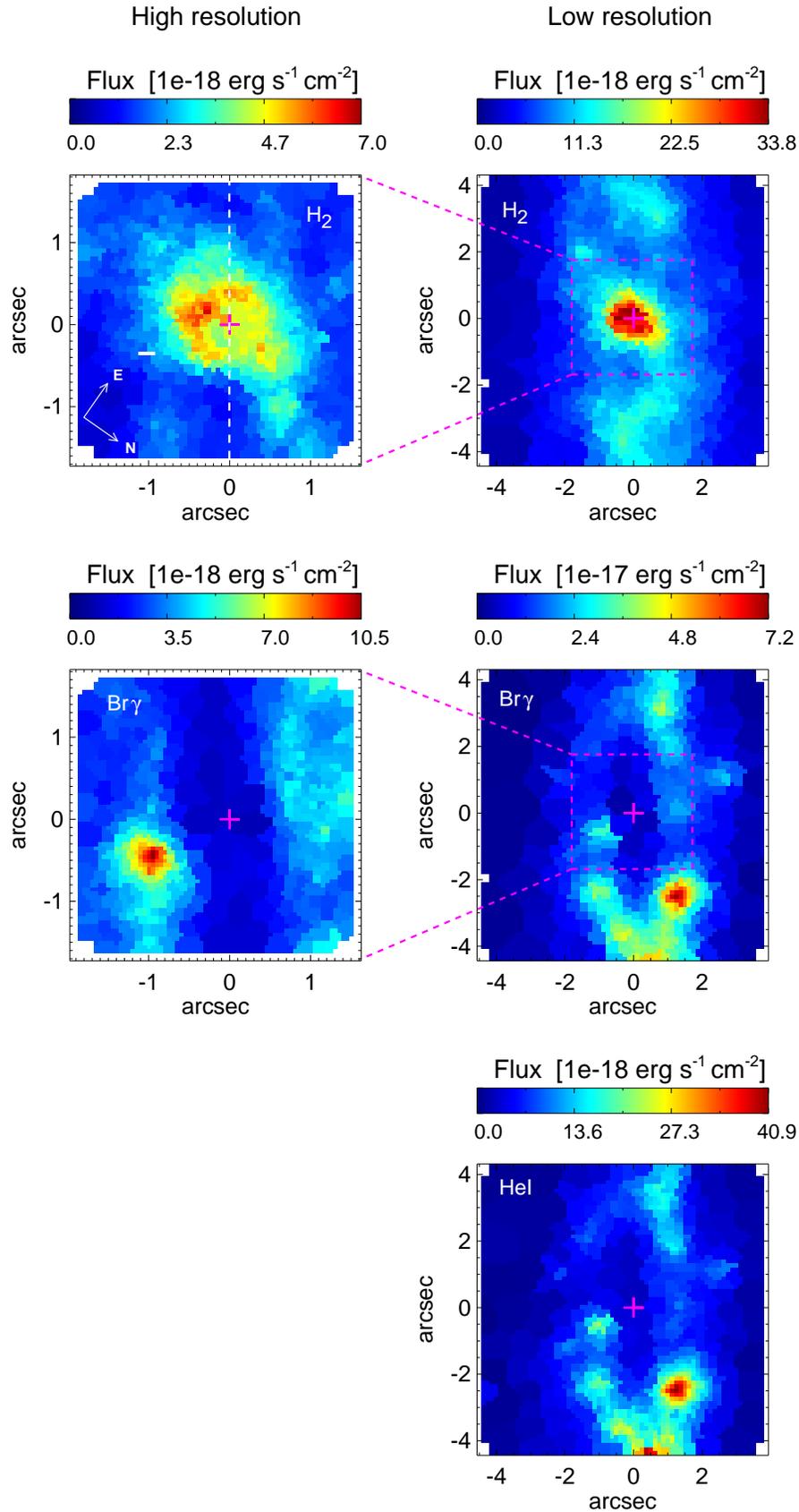}
\caption {Flux distribution maps of the \H2~2.12\micron\ (upper panels), \Brg\ (middle panels) and \HeI~2.06\micron\ (lower panel) emission lines observed in \n4536. Left panels correspond to the high-spatial resolution data and right panels to the low-spatial resolution data. The spatial orientation, indicated in the upper-left panel, is the same for all the panels. The dashed line in the upper-left panel represents the orientation of the major axis of the galaxy.}
\label{f_n4536_maps}
\end{figure*}

High- and low-resolution data of the nuclear region of \n4536 were obtained with \S, covering the inner $3'' \times 3''$ and $8''\times 8''$, respectively. Intense \Brg\ and \H2\ emission lines were detected in the spectra of this galaxy, together with weaker \HeI~2.06\micron\ emission. The flux distribution maps of \H2~2.12\micron, \Brg\ and \HeI\ derived from the \S\ data are shown in Fig.~\ref{f_n4536_maps}. The left panels correspond to the maps derived from the high-resolution data (HR maps) and the right panels show the ones derived from the low-resolution data (LR maps).

It can be seen in Fig.~\ref{f_n4536_maps} that the circumnuclear molecular and atomic gas of \n4536 exhibit very different morphologies.
The \Brg\ emission is mainly localised in a circumnuclear ring $\sim 7.5''$ (540~pc) in diameter, oriented along the photometric semi-major axis of the galaxy. The ratio between the semi-major and minor axis of the ring ($\sim 2.8$) is consistent with a circular ring with the same inclination as the galaxy (${\rm i}=69$\deg, Table~\ref{t_prop}) projected onto the sky. 
The \Brg\ emission is not uniformly distributed along the ring but shows a knotty appearance, with the strongest emission located towards the NW.
In contrast, the \H2\ emission is more centrally concentrated, with the bulk of the emission restricted to the inner $\sim 2''$ of the galaxy. The detailed structure of this region can be seen in the HR \H2\ distribution map, which shows that the \H2\ emission does not peak at the position of the active nucleus but seems to encircle it.
The nuclear region is connected to the circumnuclear ring by two ridges of weak \H2\ emission extending towards the south and the north, giving a global S-shaped appearance to the \H2\ distribution. 
In general, the morphologies of the \H2\ and \Brg\ emission-lines showed by the LR maps are consistent with those  presented by \citet{Rosenberg2012}.
The presence of strong \H2\ emission in the nucleus of \n4536 and the absence of strong hydrogen recombination lines (i.e., \Brg) support the scenario proposed by \cite{Davies1997}, in which no current star formation is taking place in the nuclear region of this galaxy (see Section~\ref{s_ionizgas} for a detailed discussion).

\n4536 is the only galaxy in the sample for which \HeI~2.06\micron\ line emission was observed.
The morphology of the \HeI\ line-emitting gas (lower-right panel of Fig.~\ref{f_n4536_maps}) has an almost perfect one-to-one correspondence with the \Brg\ emission, indicating that both \Brg\ and \HeI\ are tracing the same structures.

While previous low-resolution observations suggested the presence of a circumnuclear ring in this galaxy \citep[e.g.,][]{Vila1990, Davies1997, Laine2006}, the high spatial resolution provided by \S\ gives a clear view of the circumnuclear ring present in the inner $\sim 8''$ of \n4536.
A more detailed study of the ionized gas ring of this galaxy is presented in Section~\ref{s_ring}.

\subsection{\n4569}\label{s_n4569}

\begin{figure}
\centering
\includegraphics[width=0.75\columnwidth]{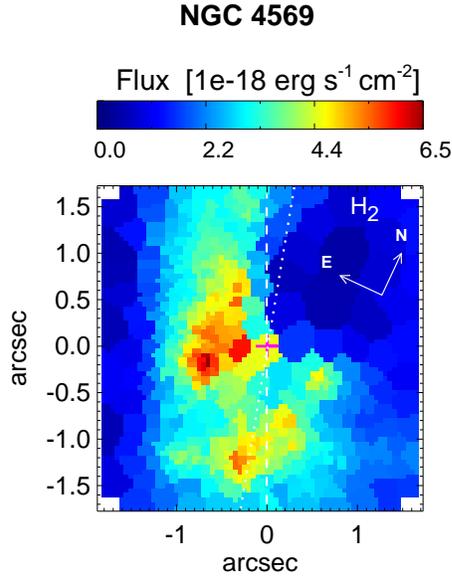}
\caption {Flux distribution map of the \H2~2.12\micron\ emission line observed in \n4569. The dashed line indicates the position of the major axis of the galaxy and the dotted line corresponds to the stellar bar position angle.}
\label{f_n4569_maps}
\end{figure}

\n4569 (M90) is a member of the Virgo cluster classified as a SAB(rs)ab galaxy.
Its nucleus was classified as a `transition object' or weak-[O\,{\sc i}] LINER by \cite{Ho1997}, although the mechanism producing the nuclear activity is still unclear, with several lines of evidence indicating a strong contribution from a nuclear starburst \citep[e.g.,][]{Keel1996, Ho1997, Maoz1998, Barth2000, Alonso-Herrero2000, Ho2001, Gabel2002}.
This galaxy has a bright, point-like nucleus at optical and $UV$ bands that dominates the emission. The HST $V-I$ map show that while the circumnuclear region is dusty, the nucleus itself is apparently unobscured by dust \citep{Pogge2000}.
\n4569 is considered an `anemic spiral' because of its \HI\ deficiency (it has $\sim 1/10$ of the atomic gas that similar field galaxies have) and small extent compared to its optical size \citep[e.g.,][]{vandenBergh1976, Cayatte1994, Koopmann2004}.
Its molecular CO gas has been extensively studied \citep[e.g.,][]{Kenney1986, Sakamoto1999a, Sakamoto1999b, Sofue2003, Helfer2003, Jogee2005, Nakanishi2005, Boone2007, Boone2011, Wilson2009}. The CO emission is highly concentrated in a compact bar-like region (of $\sim 17'' \times 6''$ extending along ${\rm P.A.} \approx 24$\deg) with three off-centre peaks, one located at $\sim 1''$ towards the north-east of the nucleus and two symmetrical peaks separated by $\sim 10''$ \citep{Boone2007}.

The \S\ observations of \n4569 were done in the high-resolution configuration, covering a total FOV of $3'' \times 3''$. The $K$-band emission line spectrum of this galaxy is dominated by \H2\ transitions, with no signs of emission from ionized gas. This is consistent with previous long-slit observations where no \Pab\ and/or \Brg\ emission was detected in the nucleus \citep[e.g.,][]{Alonso-Herrero2000, Rhee2005}. 
Fig.~\ref{f_n4569_maps} shows the \H2\ flux-distribution map derived from \H2~2.12\micron. It can be seen that the \H2\ emission line gas distribution is very irregular, with most of the emission located off-nucleus on the eastern part of the FOV. There is a rapid drop in the signal-to-noise as one goes from the eastern-southern regions towards the north, where the intensity of the \H2\ line falls below the continuum noise level.
Although the \S\ data presented here is mapping only a small part of the molecular bar observed in \n4569, it is possible to say that the general \H2\ distribution displayed by this galaxy is consistent with the CO(1--0) and CO(2--1) maps reported by \cite{Boone2007}, in the sense that the stronger emission is located towards the east and south. However, the \H2\ emission peaks closer to the nucleus than the CO emission, suggesting a strong gradient in the molecular gas temperature going from the centre to the outer regions of the galaxy.

\subsection{\n4579}\label{s_4579}

\begin{figure}
\centering
\includegraphics[width=0.75\columnwidth]{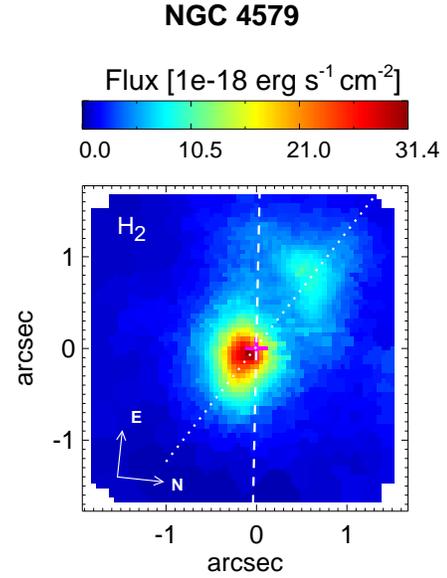}
\caption {Flux distribution map of the \H2~2.12\micron\ emission line observed in \n4579. The dashed line indicates the position of the major axis of the galaxy and the dotted line corresponds to the stellar bar position angle.}
\label{f_n4579_maps}
\end{figure}

\n4579 is an SAB(rs)b galaxy with a LINER/Seyfert 1.9 nucleus \citep{Ho1997}. As in the case of \n4569, this galaxy is considered an `anemic spiral' of the Virgo cluster. HST UV, \OIII\ and \Ha\ images show a bright nuclear point source surrounded by filamentary structures, with the brightest emission located at distances $\lesssim 2''$ towards the NE \citep{Maoz1995, Pogge2000}.
High-resolution CO intensity maps presented by \cite{Garc'ia-Burillo2005, Garc'ia-Burillo2009} showed that most of the cold molecular gas in \n4579 is located in a two spiral arm structure that can be traced from ${\rm r}\sim 10''$ down to a distance of 2\arcsec\ from the nucleus. The spiral arms lie at the leading edges of the $\sim 12$~kpc-diameter bar identified in $K$-band images \citep[e.g.,][]{Knapen2003} and are well correlated with dust lanes revealed by $V-I$ HST images \citep{Pogge2000}, especially the arm located towards the north. Some emission from CO is observed at smaller distances from the centre but not on the position of the AGN it self.

The \S\ $K$-band spectra of the inner $3'' \times 3''$ of \n4579 display strong \H2\ emission-lines and no evidence of ionized gas. Two main structures are seen in the \H2\ emission-line map shown in Fig.~\ref{f_n4579_maps}. This emission is dominated by a symmetric nuclear component, slightly displaced towards the S with respect to the maximum of the continuum. Towards the NE, at a distance of $\sim 1''$ from the nucleus, a weaker component is present.
The overall shape of the \H2\ emission resembles the one revealed by \Ha\ and \OIII\ narrow-band images, although the latter show much more complex structures resolved into individual knots \citep[see figure~1d and 3a of][]{Pogge2000}. Additionally, there is no correspondence between the \H2\ emission presented here with dust or CO emission, which is  mostly found at larger distances from the nucleus.

\section{Physical conditions of the nuclear regions}\label{s_pconditions}

Rotational and vibrational \H2\ transitions are important cooling channels for warm molecular gas, observed as a series of IR emission lines.
\H2\ emission lines are commonly observed in the NIR spectra of galaxies harbouring energetic processes related to massive star formation, shocks or AGNs, and have proved to be important diagnostic tools in all molecular media.

Most of the galaxies in our sample show signs of AGN activity in their centres and, at the same time, regions where star formation is taking place. Therefore, the data presented here give us an excellent opportunity to study the physical conditions of the \H2\ emission-line gas in these two distinct environments.
In order to achieve this we extracted spatially integrated 1D spectra from the nuclear and circumnuclear regions of the galaxies. The size and location of these regions vary from galaxy to galaxy depending on their particular morphology and are indicated in the first column of Table~\ref{t_mass}. Many emission lines were detected in the \S\ spectra of the sample galaxies, including \Brg\ and \H2~1.96, 2.03, 2.12, 2.22 and 2.25\micron. As described below, these lines provided us with a variety of tracers of different physical quantities. The emission-line fluxes of the individual lines were obtained by fitting a single Gaussian component plus a linear continuum to the spectra. The `Liner' routine \citep{Pogge1993} was used for this purpose. The measured emission-line fluxes can be found in Table~\ref{t_mass}. The precise position and size of the apertures used to extract the 1D spectra is given in Table~\ref{t_regions} of Appendix~\ref{appendix}.

\begin{figure*}
\centering
\includegraphics[width=0.75\textwidth]{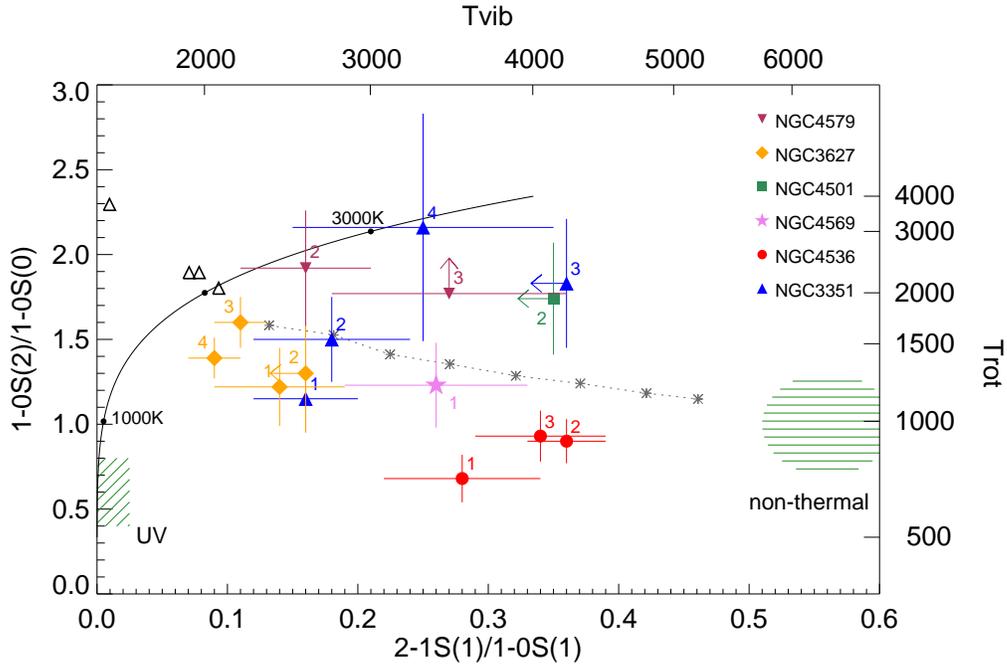}
\caption {\H2\ 2--1S(1)~2.25\micron/1--0S(1)~2.12\micron\ and 1--0S(2)~2.03\micron/1--0S(0)~2.22\micron\ emission line ratios measured from different regions of the galaxies in the sample. The code number of each measurement corresponds to the region of the galaxy specified in Table~\ref{t_mass}. The solid curve represents thermal emission in the temperature range 500--4000~K (the dots on the curve indicate T=1000, 2000 and 3000~K). Open triangles correspond to X-ray-heating models of \citet{Lepp1983} and \citet{Draine1990}. Areas of the diagram corresponding to thermal UV excitation models \citet{Sternberg1989} and non-thermal UV fluorescence models \citet{Black1987} are indicated with shaded regions. The dashed-line represents the predicted line ratios from a mix of thermal and UV-fluorescence models of \citet{Black1987}, where the percentage of the thermal component decreases in steps of 10 per cent starting at the left with a model where 90 per cent is thermal and 10 per cent is UV fluorescence. Additionally, the corresponding T$_{\rm vib}$ and T$_{\rm rot}$ temperature scales are indicated in the upper and right axes, respectively.
}
\label{f_h2_models}
\end{figure*}

\subsection{Excitation mechanisms and temperature of the \H2\ emission-line gas}\label{s_excitation}

Several mechanisms have been proposed for the excitation of the ground state ro-vibrational levels of \H2. These can be separated into two main groups: thermal and non-thermal processes. 
In thermal processes, the \H2\ molecule is excited via inelastic collisions with atoms or molecules in warm (T$\sim 500-2000$~K) gas, heated either by shocks or UV/X-ray radiation in high density regions \citep[e.g.,][]{ Lepp1983, Sternberg1989, Brand1989, Draine1990, Maloney1996}. In this case, to a first approximation, the \H2\ line-emitting gas is in local thermal equilibrium (LTE) and the spectrum is therefore characterized by a single excitation temperature. Since \H2\ is rapidly dissociated by thermal collisions above $\sim 4000$~K, the excitation temperature of the thermal component must be lower than this. 
In non-thermal processes, the NIR ro-vibrational lines are observed due to a radiative cascade through the excited vibrational and rotational levels of the ground electronic state. The excitation of the electronic state can be produced either by near-UV (11--13.6~eV) pumping to the electronically excited Lyman and Werner bands \citep[e.g.,][]{Black1987} or by a collision with a fast electron produced by X-ray ionization to singlet electronic states \citep[e.g.,][]{ Gredel1995}.

Each of the excitation mechanisms mentioned above produces different \H2\ spectra, which are often observed together within a single object \citep[e.g.,][]{Mouri1994, Veilleux1997, Davies1997, Davies2005, Quillen1999, Rodr'iguez-Ardila2004, Rodr'iguez-Ardila2005b, Riffel2006a, Riffel2008, Riffel2010, Zuther2007, Storchi-Bergmann2009}. The relative intensity of the \H2\ emission lines can be used to infer the dominant process responsible for the \H2\ emission. In particular, the \H2\ 2--1S(1)~2.25\micron/1--0S(1)~2.12\micron\ line ratio is an excellent discriminator between non-thermal and thermal processes, while 1--0S(2)~2.03\micron/1--0S(0)~2.22\micron\ can be used to differentiate between thermal UV excitation and X-ray/shock excitation \citep{Mouri1994}. Additionally, these line ratios are sensitive to the vibrational temperature (T$_{\rm vib}$) and rotational temperature (T$_{\rm rot}$) of the emission-line gas, respectively \citep[e.g.,][]{Scoville1982, Reunanen2002}. The excitation temperature (T$_{\rm vib}$ or T$_{\rm rot}$, depending on the transitions involved) will depend on the relative importance of the radiative and collisional processes. In the case of gas in LTE the ro-vibrational level populations are characterized by a thermalised Boltzman population and the excitation temperature corresponds to the kinetic temperature of the gas.
Assuming the computed A-Einstein coefficients of \cite{Turner1977}, the relation between the excitation temperature of the gas and the \H2\  2--1S(1)/1--0S(1) and 1--0S(2)/1--0S(0) emission line ratios can be expressed as:

\begin{equation}\label{eq1}
 {\rm T}_{\rm vib} \simeq {5600 \over {{\rm ln}~ (1.355 \times {I_{1-0S(1)} \over I_{2-1S(1)}}) }},
\end{equation}
\begin{equation}\label{eq2}
{\rm T}_{\rm rot} \simeq {-1113 \over {{\rm ln}~ (0.323 \times{ I_{1-0S(2)} \over {I_{1-0S(0)}}}) }}.
\end{equation} 
Fig.~\ref{f_h2_models} shows the \H2\  2--1S(1)/1--0S(1) vs. 1--0S(2)/1--0S(0) line-ratio diagnostic diagram for the selected regions of the galaxies in our sample, together with diverse key excitation models taken from the literature. Each filled symbol represents one particular galaxy and the associated numbers indicate the different regions specified in the first column of Table~\ref{t_mass}.
The right and upper axes show the values of T$_{\rm vib}$ and T$_{\rm rot}$, respectively, derived from Eqs.~(\ref{eq1}) and (\ref{eq2}) as a function of the two line ratios. Table~\ref{t_mass} lists the temperature values obtained for the different regions of the galaxies. 
From this figure it is clear that there is no single excitation mechanism acting in the nuclear and circumnuclear regions of the galaxies, regardless of their type of activity. In general, each region is characterized by a T$_{\rm rot}$ lower than its T$_{\rm vib}$, consistent with some contribution from non-thermal processes. The values of temperature derived vary from $\sim 700$ up to 4000~K, with a tendency for the objects to be located near the pure thermal excitation curve for the range of temperatures $\sim 1000-3000$~K. These high temperatures  correspond to the region on the diagram where X-ray/shock thermal excitation dominates. 
\cite{Dors2012} constructed a grid of photoionization models using the code {\sc cloudy}/08 \citep{Ferland1998}, which includes both collisional and radiative interactions of \H2\ and its environment \citep{Shaw2005}. The input parameters are a spectral energy distribution of the ionizing source similar to that observed in typical AGNs, three values of metallicity (0.5, 1 and 2 solar metallicy), an electronic density $n_e=10^4$~cm$^{-3}$ and an ionization parameter $U$ in the $-4.0\le {\rm log} U \le -1$ range. The resulting \H2\ line ratios span a large range of values, \H2~$2.25/2.12~\mu{\rm m}\sim 0.05-0.5$ and $2.03/2.22~\mu{\rm m}\sim 1.1-2.0$, reproducing most the observed \H2\ line ratios of the galaxies in our sample.

While most of the galaxies in our sample are characterized by a value of \H2~2.03/2.22~\micron\ higher than 1, this ratio is lower in the case of \n4536. The \H2\ line ratios of this galaxy occupy a region on the diagram dominated by a mix of UV thermal and non-thermal models, suggesting that UV radiation plays a key role in the excitation of the \H2\ in this galaxy, with no significant contribution from shocks and/or X-ray radiation.
The combination of low T$_{\rm rot}$ and high T$_{\rm vib}$ make this galaxy unique among the objects presented here, placing it in a region of the diagram typically occupied by photodissociation regions.

In summary, the analysis presented in this section suggest that thermal processes are primarily responsible for the excitation of  the \H2\ gas in the nuclear and circumnuclear regions of the galaxies in our sample, with some minor contribution from non-thermal processes (i.e., UV fluorescence).

\begin{table*}
\caption{Emission line fluxes (in units of $10^{-16}$~erg~cm$^{-2}$~s$^{-1}$), temperatures, emission-line ratios and molecular masses measured for different regions of the galaxies of the sample. The errors correspond to $3\sigma$.}
\label{t_mass}
\begin{tabular}{lcccccccccc}
\hline
\multirow{2}*{Galaxy}  &\H2\ 1--0S(2) & \H2\ 1--0S(1) & \H2\ 1--0S(0) & \H2\ 2--1S(1) & \Brg\ & T$_{\rm vib}$ & T$_{\rm rot}$ & \multirow{2}*{$\frac{{\rm H}_2~2.12~\mu{\rm m}}{{\rm Br}\gamma}$} & M$_{\rm warm}$ & M$_{\rm cold}$ \\
	& 2.03\micron & 2.12\micron & 2.22\micron & 2.25\micron & 2.16\micron  & [K] & [K] & &  [M$_{\odot}$]  & [$10^6 {\rm  M}_{\odot}$]\\
\hline
\n3351 &&&&&&&&&\\
\hspace{0.2cm} 1 -- Nucleus 		&$3.9\pm1.3$&$7.5\pm1.6$&$3.4\pm1.2$&$1.2\pm0.9$&$1.3\pm0.9$& 2600 & 1125 & 5.9 & 3.9  & 2.8  \\
\hspace{0.2cm} 2 -- W cloud 		&$0.7\pm0.1$&$1.2\pm0.2$&$0.5\pm0.2$&$0.2\pm0.1$&--& 2800 & 1500 & -- & 0.6  & 0.4  \\
\hspace{0.2cm} 3 -- $3'' \times 3''$ 	&$21.4\pm5.8$&$16.7\pm6.7$&$11.9\pm6.7$&$<6.0$&--& $<4300$ & 2000 & -- & 8.7  & 6.2    \\
\hspace{0.2cm} 4 -- $8'' \times 8''$ 	&$117\pm55$&$151\pm26$&$54.3\pm28.7$&$37.5\pm28.1$&$150\pm25$& 3300 & 3000 & 1.0 & 79.0   & 57.0   \\
\hspace{0.2cm} 5 -- Ring 	        &--&$89.0\pm22.4$&--&--&$160\pm28$& -- & -- & 0.56 &  46.5 &  33.3   \\
\n3627 &&&&&&&&&&\\
\hspace{0.2cm} 1 -- Nucleus 		&$18.2\pm6.9$&$43.1\pm6.0$&$15.0\pm5.7$&$6.2\pm5.7$&$20.6\pm6.6$& 2500 & 1200 & 2.1 & 22.3 & 16.0  \\
\hspace{0.2cm} 2 -- $3'' \times 3''$ 	&$41.8\pm22.9$&$108\pm18$&$32.1\pm17.4$&$<17.0$&--&  $<2600$ & 1300 & --  & 55.6 & 40.0 \\
\hspace{0.2cm} 3 -- N cloud             &$4.7\pm0.9$&$11.7\pm0.5$&$3.0\pm0.6$&$1.3\pm0.6$&--& 2207 & 1680  & --  & 6.0  & 4.4  \\
\hspace{0.2cm} 4 -- S cloud             &$6.0\pm0.8$&$15.2\pm0.9$&$4.2\pm0.9$&$1.4\pm1.0$&--& 2091 & 1388  & --  & 7.8  & 5.6  \\
\hspace{0.2cm} 5 -- $8'' \times 15''$ 	&--&$268\pm62$&--&--&--&  -- & -- & -- & 138 & 100  \\
\n4501 &&&&&&&&&&\\
\hspace{0.2cm} 1 -- Nucleus 		&--&$7.6\pm3.9$&$5.8\pm4.5$&--&--& -- & -- & --  & 10.5 & 7.5  \\
\hspace{0.2cm} 2 -- Ring		&$48.1\pm27.0$&$50.1\pm12.9$&$27.1\pm18.0$&$<18.0$&--& $< 4200$ & 2000 & --  & 69.4 & 49.9  \\
\hspace{0.2cm} 3 -- $3'' \times 3''$    &--&$65.4\pm20.3$&--&--&--&  --    & --    & -- & 90.6 & 65.0    \\
\n4536 &&&&&&&&&&\\
\hspace{0.2cm} 1 -- Nucleus 		&$11.5\pm6.2$&$32.6\pm5.6$&$17.1\pm5.9$&$9.0\pm5.9$&$17.0\pm5.2$& 3520 & 730 & 1.9  & 37.0  & 26.6   \\
\hspace{0.2cm} 2 -- Ring 	 	&$92.2\pm26.0$&$240\pm29$&$105\pm22$&$88\pm25$&$532\pm29$& 4300 & 885 & 0.45 & 272.6 & 196.0   \\
\hspace{0.2cm} 3 -- $8'' \times 8''$ 	&$140\pm59$&$331\pm34$&$151\pm36$&$114\pm36$&$567\pm33$& 4000 & 920 & 0.6 & 375.9 & 270.0  \\
\n4569 &&&&&&&&&&\\
\hspace{0.2cm} 1 -- $3'' \times 3''$ 	&$52.2\pm21.2$&$95.4\pm18.8$&$42.5\pm19.5$&$24.8\pm19.5$&--& 3400 & 1200 & --& 74.1  & 53.3    \\
\n4579 &&&&&&&&&\\
\hspace{0.2cm} 1 -- Nucleus 		&$29.5\pm4.5$&$46.8\pm7.5$&--&--&--&  -- & -- & -- & 64.8 & 46.6 \\
\hspace{0.2cm} 2 -- NE cloud 		&$10.2\pm2.3$&$19.0\pm2.4$&$5.3\pm2.5$&$3.0\pm2.4$&--& 2600 & 2300 & -- & 26.4  & 19.0    \\
\hspace{0.2cm} 3 -- $3'' \times 3''$ 	&$61.8\pm25.8$&$82.9\pm21.3$&$34.9\pm19.7$&$<22.5$&--& 3500 & $> 2000$ & -- & 114.9 & 82.6  \\

\hline
\end{tabular}
\end{table*}

\subsection{Ionized gas and the \H2/\Brg\ emission line ratio}\label{s_ionizgas}

The \Brg\ recombination line was observed in half of the galaxies of the sample and, even though the number of objects is low, we can appreciate very different morphologies displayed by this line. In two cases (\n3351 and \n4536) the flux distribution maps of \Brg\ show a clear ring-like morphology, while in the third case (\n3627) the bulk of emission is restricted to a relatively compact nuclear region.

In general, the \Brg\ distribution of the galaxies in the sample does not follow their \H2\ emission. 
A clear deficit of \Brg\ emission relative to the \H2\ emission is observed in the nuclei of \n3351 and \n4536. In order to quantify this, we included in Table~\ref{t_mass} the values of the \H2~2.12\micron/\Brg\ emission line ratio measured for the different regions of the galaxies. 
These values vary between 0.45 and 5.9, in agreement with the usual values found for starburst galaxies and AGNs \citep[e.g.,][]{Moorwood1988, Moorwood1990, Larkin1998, Rodr'iguez-Ardila2004, Rodr'iguez-Ardila2005b, Riffel2010}. The expected line ratio for starburst--dominated galaxies is typically low ($\lesssim 0.6$), while LINER nuclei show the highest ratios, usually $\gtrsim 2$. Composite and Seyfert nuclei are usually characterized by a \H2/\Brg\ ratio of order unity. 
The lowest of our values are the ones measured in the circumnuclear regions and agree with the expected values for pure starburst emission. On the other hand, the highest values are the ones measured in the nucleus of the galaxies, which are in the range usually occupied by LINER nuclei. While this is consistent with the nuclear classification of \n3627 and \n4536, the galaxy \n3351 is known to be a non--AGN. However, the high value of \H2/\Brg\ displayed by the nucleus of this galaxy can be explained as a consequence of the relative old stellar population that characterizes its innermost regions (see below), in which the presence young OB stars responsible for the \Brg\ emission has declined and the number of SN explosions, which heat the gas via shocks and produce the \H2\ emission, has increased. 
Although a detailed interpretation of the variation of the \H2/\Brg\ from pure--starburst emission to Seyfert and LINER nuclei is non-trivial, one appealing scenario is that the different line ratios reflect different excitation mechanisms, in the sense that they describe a transition from purely ionizing radiation powered by star formation to pure shock excitation driven by SNRs.
The \H2/\Brg\ derived for the galaxies together with the analysis of the excitation mechanisms presented in the previous section are compatible with this picture: the lowest \H2/\Brg\ ratios are displayed by the outer regions of \n4536, which are the ones with the strongest non-thermal contribution (UV fluorescence) to the excitation of the \H2\ molecule, while the highest \H2/\Brg\ ratios are measured for regions with a strong influence from thermal processes
(e.g., shock heating).

\subsubsection{Star--forming rings in \n3351 and \n4536}\label{s_ring}

The \S\ data presented here allowed us to spatially resolve the innermost regions of \n3351 and \n4536, both of which  show circumnuclear rings traced by ionized gas (i.e. \Brg\ and/or \HeI\ emission lines).
This type of ring is frequently found in early to intermediate Hubble--type barred galaxies, and is usually associated with inflowing gas accumulating near the inner Lindblad resonances of bars \citep[see][and references therein]{ Buta1996}.
Photoionization by OB stars is the most likely source of the nebular emission, although shocks produced by stellar winds and supernovae (related to the same stellar population) may play an additional role.
In addition, high energy photons from AGN can also ionize the ISM and contribute to the \Brg\ emission. Although \n4536 is known to harbour a low-luminosity AGN, the spatial distribution of its \Brg\ emission (Fig.~\ref{f_n4536_maps}) indicates that this contribution is relatively low and restricted to the innermost regions, with no or negligible influence in the circumnuclear ring observed in this galaxy. Therefore, outside the nucleus we consider \Brg\ as a primary tracer of young stellar populations.

It is possible to estimate upper limits of the age of the star formation occurring in these galaxies by measuring the equivalent width of the \Brg\ line, W$_{\rm Br\gamma}$. For this, we extracted two spectra for each galaxy, one integrating the emission of the inner $\sim 1''$ of the galaxies and a second one encompassing the emission coming from the ring. Note that in the case of \n3351 only part of the circumnuclear ring is covered by the field of view of our data (Section~\ref{s_3351}).
We measured W$_{\rm Br\gamma}$ of 0.3 and 2.1~\AA\ for \n3351 and 0.8 and 8.3~\AA\ for \n4536, with the higher values corresponding to the ring spectra. 
Using the Starburst99 models \citep{Leitherer1999} for an instantaneous burst, different slopes for the initial mass function (2.35 and 3.33), upper mass limits of 30 and 100\Ms\ and solar metallicity, the 
measured values of W$_{\rm Br\gamma}$ predict ages of the stellar populations of 10.5 and 6.9~Myr for \n3351 and 7.2 and 6.5~Myr for \n4536, with the younger populations located in the circumnuclear rings. 
As mentioned above, the age values correspond to upper limits, since we have not taken into account the possible contribution from the AGN and/or underlying (old) stellar population to the continuum. This additional component increases the continuum level, producing lower values of W$_{\rm Br\gamma}$ than the true ones and, hence, larger ages than the expected for pure star--forming regions. 
An estimation of the contribution of the AGN to the continuum can be made based on the values derived for the equivalent width of the CO(2--0) absorption line at 2.29\micron, W$_{\rm CO(2-0)}$, as proposed by \citet{Davies2007b}. These authors showed that the intrinsic W$_{\rm CO(2-0)}$ of any stellar population is typically $\sim 12$~\AA\ (with an uncertainty of $\pm 20$ per cent), independent of the star formation history and age. Therefore, for a stellar continuum diluted by additional non-stellar emission, the fraction of the stellar light can be estimated by the ratio of the observed and intrinsic equivalent widths of the CO feature. In order to check for a possible AGN contribution to the continuum in \n4536 we measured the W$_{\rm CO(2-0)}$ in the nuclear and ring spectra of this galaxy. The values are W$_{\rm CO(2-0)} = (13.6 \pm 0.8)$~\AA\ for the nuclear region and $(14.0 \pm 0.7)$~\AA\ for the ring. These values are in the range derived by \citet{Davies2007b}, suggesting that the contribution from the AGN to the observed continuum in \n4536 is negligible.

Alternatively, we could assume a scenario where continuous star formation is taking place. In this case, the Starburst99 models predict ages of more than 60~Myr for the measured W$_{\rm Br\gamma}$. Such old ages are improbable, since the star formation would need to be sustained over long periods of time in very localized regions, in which turbulence and heating from SNe will disrupt the molecular clouds after $\sim 20-30$~Myr, inhibiting further star formation \citep[e.g.,][]{Blitz2007}.

In addition to \Brg, the spectra of \n4536 also display \HeI~2.06\micron\ emission along the circumnuclear ring (Fig.~\ref{f_n4536_maps}).
The ionization energy required to produce \HeI\ recombination lines is 24.6~eV, almost a factor of two higher than that of the hydrogen (13.6~eV). Therefore, \HeI\ emission lines are expected to arise in the vicinity of very hot massive stars (most massive OB stars) and AGNs, where enough high-energy photons are available to ionize the \HeI\ gas. Since the hottest stars will vanish fastest, the \HeI/\Brg\ emission-line ratio can give us a qualitative estimation of the relative ages of the young star clusters located along the circumnuclear ring observed in \n4536.
We measured a value of \HeI/\Brg\ $\simeq 0.5$ along the entire ring, showing no significant variations and/or gradients. Hence, the young star clusters along the ring of this galaxy have the same ages, suggesting a scenario where the star formation started at the same time over the entire structure.
It must be kept in mind that this is only a qualitative interpretation, since a detailed assessment of the \HeI/\Brg\ ratio is complicated by additional dependences such as the geometry of the ionized gas, its density, dust content and helium abundances \citep[e.g.,][]{Shields1993, Lumsden2001}.

\section{Nuclear mass inventory}\label{s_mass}

\subsection{Warm molecular gas mass}\label{s_warm_mass}

The total mass of the warm molecular hydrogen, M$_{\rm warm}$, can be derived from the observed flux of the \H2\ 1--0S(1) 2.12\micron\ emission line. Assuming a temperature ${\rm T}=2000$~K,  the \H2\ 1-0S(1) transition probability A$_{\rm S(1)}=3.47\times 10^{-7}$~s$^{-1}$, and a population fraction in the $(\nu,J)=(1,3)$ level $f_{(1,3)}=0.0122$, the relation between the warm molecular mass and the \H2\ 1-0S(1) emission line flux is given by:

\begin{equation}\label{eqMwarm}
 {\rm M}_{\rm warm} \simeq {5.0875 \times 10^{13}
\left(\frac{D}{{\rm Mpc}} \right)^2  \left( \frac{{\rm F}_{\rm 1-0S(1)}}{{\rm erg~s^{-1}cm^{-2}}}\right) 10^{0.4~A_{2.2}}},
\end{equation} 
where $D$ is the distance to the galaxy and $A_{2.2}$ is the extinction at 2.2\micron\ \citep[e.g.,][]{Scoville1982}. Unfortunately, the data presented here do not allow us to determine the extinction affecting the inner regions of the galaxies in our sample and therefore it is only possible to determine a lower limit for \Mw, assuming a value of $A_{2.2}=0$. Taking into account that the absorption at the $K$-band is only $\approx10$ per cent of that at visible wavelengths \citep{Cardelli1989}, and a typical extinction $A_V \sim 1.2$~mag \citep{Ho1997}, correcting for internal extinction will not increase the derived masses by more than 12 per cent (or a factor of 2 for an extreme case of $A_V\sim 7.7$). 
Table~\ref{t_mass} lists the integrated masses of warm molecular gas derived for the different regions of the sample galaxies studied. These cover a wide range of values, ranging from $\sim 9$\Ms\ in the inner 75~pc of \n3351 up to $\sim 300$\Ms\ in the inner 280~pc of \n4536.

\subsection{Cold molecular gas mass}\label{s_cold_mass}

In principle, the \H2\ 1--0S(1) 2.12\micron\ emission does not necessarily reflect the cold gas distribution, since it also depends on an external source of energy capable of exciting the \H2\ ro-vibrational levels, such as UV photons from young OB stars, shocks driven by SNRs or by outflows or winds from the active nucleus (see Sect.~\ref{s_excitation}).
Historically, the favoured proxy for estimating the amount of cold \H2\ gas has been CO emission.
However, it has been suggested that an estimate of the total content of cold \H2\ gas can be obtained from the \H2~2.12\micron\ line luminosity (L$_{\rm 1-0S(1)}$).
This method has the advantage of being able to probe the inner-most regions of nearby objects with higher spatial resolution than CO observations. 
\cite{Muller-S'anchez2006} derived a mean conversion factor of 4000\Ms/\Ls\ (with a standard deviation of a factor of $\sim 2$) from the relation between the \H2~2.12\micron\ luminosity and the \H2\ mass traced by CO luminosity in a sample of sixteen luminous and ultraluminous infrared galaxies\footnote{One outlier, \n6240, with a corresponding factor of only 400\Ms/\Ls, was taken out of the sample.}. Taking into account Eq.~(\ref{eqMwarm}), these values yield cold-to-warm \H2\ mass ratios in the range of M$_{\rm cold}$/M$_{\rm warm} \simeq (1-5) \times 10^6$.
On the other hand, \cite{Dale2005} studied a large sample of active and star forming galaxies and found cold-to-warm \H2\ mass ratios covering a much larger range of values, M$_{\rm cold}$/M$_{\rm warm} \simeq 10^{5-7}$.

\begin{figure}
\includegraphics[width=\columnwidth]{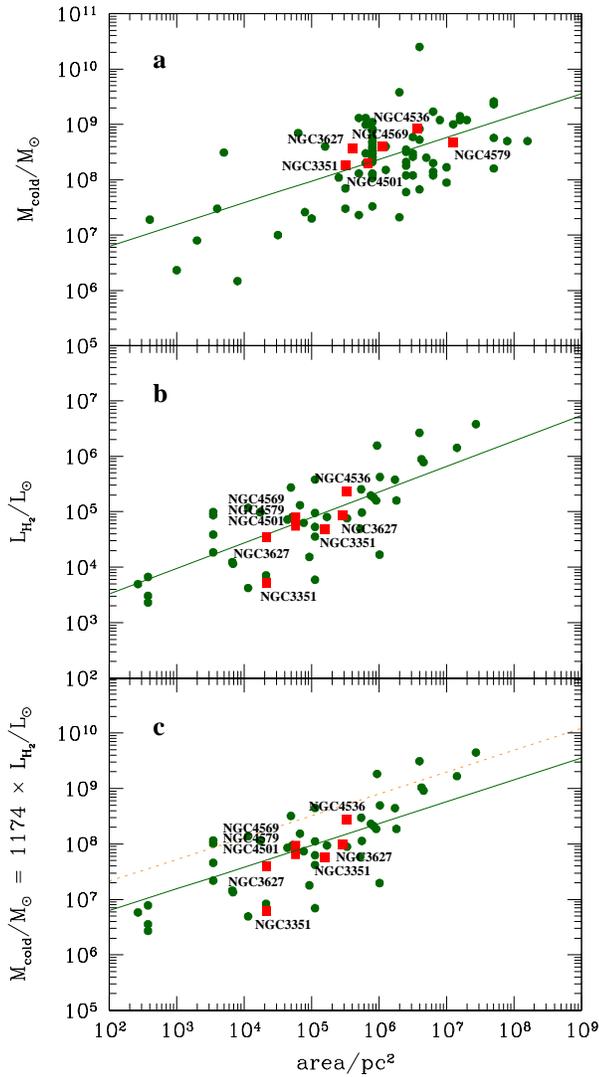}
\caption{Panel (a): Cold \H2\ gas masses derived from CO observations as a function of the integrated area. The solid line corresponds to the power-law that best reproduces the data.
Panel (b): \H2~2.12\micron\ luminosities as a function of the integrated area. The solid line corresponds to a power-law fit to these data.
Panel (c): Cold \H2\ gas masses derived from \H2~2.12\micron\ luminosities using a rescaling factor $\beta = 1174$\Ms/\Ls. The solid line corresponds to the same power-law as in panel (a), while the dashed line corresponds to the relation that would be obtained using a factor of $\beta =4000$\Ms/\Ls\ (see the text for details).
Red squares correspond to values of the galaxies in our sample.}
\label{f_Mcold}
\end{figure}

In order to investigate the relation between the \H2~2.12\micron\ emission and the cold \H2\ mass, we compiled from the literature values of M$_{\rm cold}$ derived from CO observations and \H2~2.12\micron\ luminosities\footnote{References for the CO data: \cite{Sakamoto1999a, Sakamoto1999b, Sakamoto2011, Walter2001, Walter2002, Schinnerer2002, Garc'ia-Burillo2003, Gao2004, Combes2004, Combes2009, Onodera2004, Garc'ia-Barreto2005, Jogee2005, Boone2007, Iono2007, Hicks2009, Israel2009, Zhu2009, Casasola2010, Casasola2011, Mart'in2010, Meier2010, Olsson2007, Olsson2010, Alatalo2011, Crocker2011, Salome2011}. References for \H2~2.12\micron\ line luminosities: \cite{Dale2004, Rodr'iguez-Ardila2004, Davies2006, Muller-S'anchez2006, Zuther2007, Boker2008, Nowak2008, Riffel2008, Riffel2010, Hicks2009, Storchi-Bergmann2009, Friedrich2010}.} for a large number of galaxies, covering a wide range of luminosities, morphological types and nuclear activity. 
Panel (a) of Fig.~\ref{f_Mcold} shows the values of M$_{\rm cold}$ derived from CO observations as a function of the integrated area (i.e., size of the aperture used to integrate the CO flux).  
The relation between the cold \H2\ gas mass and the integrated area, A, can be described as a power-law M$_{\rm cold} = \alpha {\rm A}^\gamma$, with the best-fitting parameters $\alpha = (1.0 \pm 0.3) \times 10^6$ and $\gamma = 0.39 \pm 0.06$. Many factors contribute to the large dispersion observed in this relation ($\sigma({\rm log(M)})=0.5$), including uncertainties in the CO-to-\H2\ conversion factor (which likely differs from galaxy to galaxy), the inhomogeneity in the type of galaxies included, and the intrinsic irregularity in the gas distribution of the galaxies (e.g. presence of bars, rings). 
Similarly, panel (b) of Fig.~\ref{f_Mcold} shows the distribution, as a function of the integrated area, of the \H2~2.12\micron\ luminosities compiled from the literature. These follow a similar trend as the cold \H2\ masses, with a value of $\gamma = 0.49 \pm 0.07$ consistent at $1\sigma$ with the one derived for the M$_{\rm cold}$ distribution.
This is in agreement with the previous suggestion of a proportionality between the warm and cold molecular gas masses and gives us an opportunity to obtain a rough estimation of the factor relating these two quantities. Assuming that the two relations have the same slope ($\gamma = 0.39$), the proportionality factor can be estimated by minimizing the residuals of the rescaled luminosities, $\beta \times {\rm L_{1-0S(1)}}$, with respect to the power--law that best describes the distribution of cold \H2\ masses derived from the CO measurements. 
Panel (c) of Fig.~\ref{f_Mcold} shows the values of L$_{\rm H_2}$ compiled from the literature rescaled by the factor $\beta = 1174$\Ms/\Ls\ obtained in this way. 
We also included the result obtained using the rescaling factor of 4000\Ms/\Ls\ proposed by \cite{Muller-S'anchez2006} (dashed line). Note that, if we had used that factor, we would had overestimated the masses as compared to the ones derived from the CO observations. Therefore, the factor of 1174 seems to give more representative cold \H2\ masses for a larger range of galaxy types. This factor and its corresponding error ($\sigma{\rm log }(\beta) = 0.35$) can be translated to a range of \Mc/\Mw\ ratio of $ \simeq (0.3-1.6) \times 10^6$, which is consistent with the one derived by \cite{Dale2005} and marginally consistent with the one inferred by \cite{Muller-S'anchez2006}.

From the analysis described above, we can conclude that a rough estimate of the cold \H2\ gas mass can be obtained from the integrated \H2~2.12\micron\ luminosity using the relation

\begin{equation}\label{eq_Mcold}
{\rm \frac{M_{cold}}{M_{\odot}} \approx 1174 \times \frac{L_{1-0S(1)}}{L_{\odot}}}.
\end{equation} 
Taking this relation into account, we calculated the total (cold) \H2\ gas content for the galaxies in our  sample.
The observed \H2~2.12\micron\ line fluxes and corresponding cold \H2\ masses of the nuclear and circumnuclear regions of the sample galaxies are listed in Table~\ref{t_mass}.
We found masses between $4\times 10^5$ and $2.7 \times 10^8$\Ms\ within regions with radii of up to 280~pc. In order to place these results within a larger context, we included in the lower panel of Fig.~\ref{f_Mcold} the cold \H2\ masses derived from the \H2~2.12\micron\ luminosities measured for the largest regions in each galaxy. The masses obtained for these galaxies agree very well with the `global' trend followed by the large pool of masses taken from the literature. The only galaxy that seems to deviate from this trend is \n3351, for which the inner regions show a lower cold-to-warm mass ratio than the rest of the sample. However, this ratio goes up when it is calculated within a larger radius region (i.e., $8'' \times 8''$).

The values of \Mc\ derived from the \H2\ luminosities can be compared with values derived from CO observations. For this, we also included in the upper panel of Fig.~\ref{f_Mcold} values of \Mc\ derived from CO observations for the galaxies in our sample [\n3351: \cite{Devereux1992}, \n3627: \cite{Regan2002}, \n4501: \cite{Onodera2004}, \n4536: \cite{Jogee2005}, \n4569: \cite{Boone2007}, \n4579: \cite{Garc'ia-Burillo2009}]. 
Taking into account the different sizes of the regions probed by the \H2~2.12\micron\ and CO observations, we can conclude that both measurements yield consistent results.

\subsection{Stellar and gas mass distributions and its implications for the \Mbh\ estimation}\label{s_stellar_mass}

\begin{figure*}
\centering
\includegraphics[width=\textwidth]{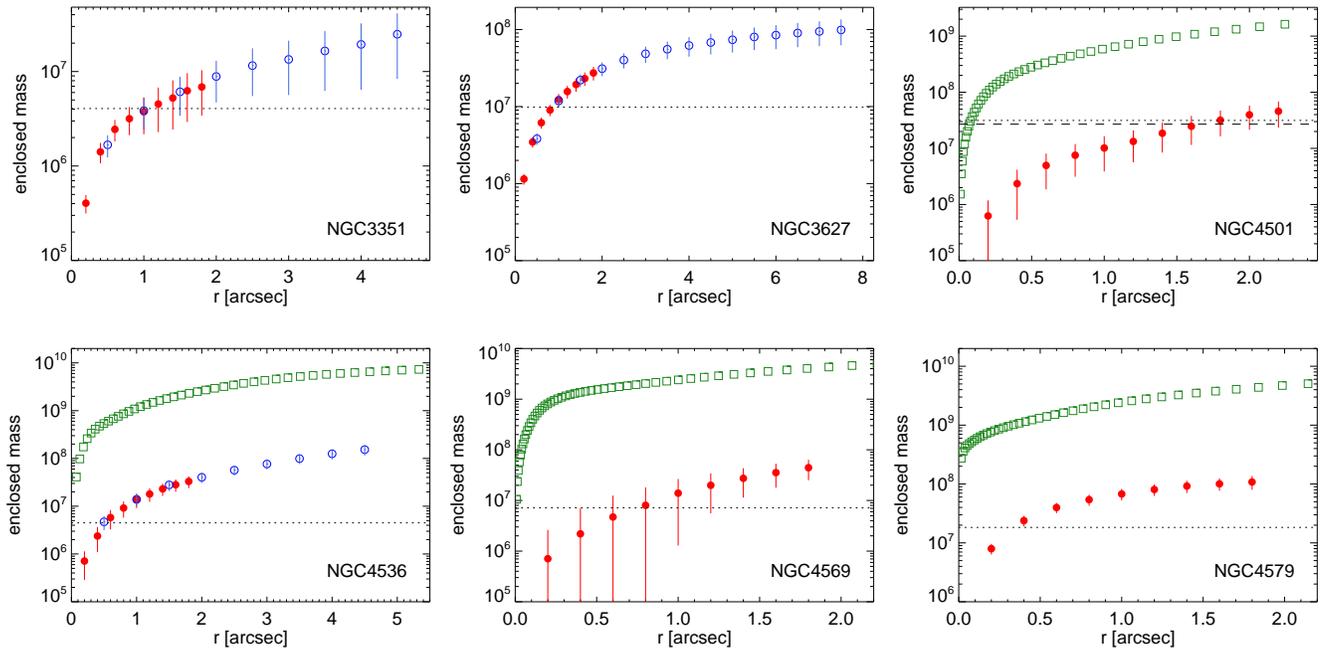}
\caption{De-projected enclosed molecular and stellar masses measured inside a given radius for the galaxies in the sample. Blue open circles and red filled circles correspond to the molecular mass measured from the low-resolution and high-resolution data cubes, respectively. Green squares indicate the enclosed stellar mass distribution derived  for \n4501, \n4536 and \n4569. The error bars reflect the 3$\sigma$ uncertainties in the \H2\ emission-line flux measurements. The dotted line in each panel indicates the \Mbh\ of the galaxy derived from its sigma (Table~\ref{t_mbh}) and the dashed line in the upper-right panel indicates the values of the \Mbh\ of \n4501 derived from the dynamical modelling.}
\label{f_enclosed}
\end{figure*}

Modelling the dynamics of stars in galactic nuclei is one of the preferred methods to probe the central potential of inactive galaxies and directly determine the mass of the central BH. However, in dynamical modelling, the {\em total} enclosed mass is constrained, including not only stars and the putative central BH, but also other components that could be present in the nucleus, such as dark matter (DM) and gas. The estimation of the relative importance of these components is essential, since it will be reflected in the derived mass-to-light ratio (M/L ratio) and \Mbh. The need for including a DM halo in the dynamical modelling has been the focus of a number of recent publications \citep[e.g.,][]{Gebhardt2009, Shen2010, McConnell2011b, Schulze2011, Gebhardt2011, Jardel2011, McConnell2012, Rusli2012a}. 
It becomes an issue if the sphere of influence of the BH is not well resolved (Rusli et al. 2012a). The omission of a separate DM component forces the DM to be incorporated into the supposed `stellar' mass. Hence, the stellar M/L ratio increases. Since M/L ratio is assumed to be radially constant this results in an overestimation of the central `stellar' mass that, in turn, is compensated by a lower \Mbh. Significant amounts of gas could affect the measured \Mbh\ similarly. If the gas adds mass outside the BH's sphere of influence its omission could, in analogy to DM, force again a higher M/L ratio and low \Mbh. If instead, the gas is concentrated near the BH, its mass could go directly in the \Mbh, biasing it too high.
Here, we will address the question of how much the presence of gas could affect the dynamical \Mbh\ measurements.

In the previous section we showed that it is possible to estimate the mass of the molecular gas from the \H2~2.12~\micron\ emission line flux. We used this result to derive the de-projected gas masses enclosed within a given radius for the galaxies in the sample. For this, we assumed that the gas is contained in a thin disc located in the plane of the galaxy and measured the \H2\ emission-line flux in spectra extracted from the continuum subtracted cubes using elliptical apertures. The size and shape of the apertures were determined by the inclination angle and semi-major axis of the corresponding galaxy (Table~\ref{t_prop}), in a way that correspond to circular apertures on the plane of the disc/galaxy. Fig.~\ref{f_enclosed} shows the de-projected enclosed masses as a function of the radius measured for the galaxies in the sample. The molecular gas masses measured inside the smallest aperture, ${\rm r}  = 0.2''$ (or 10--16~pc), vary between $4\times10^5 - 8\times10^7$\Ms, and reach $\sim 10^8$\Ms\ inside larger apertures. Particularly noticeable is the relatively high gas mass content of the galaxy \n4579, which shows the highest measured mass value of the whole sample inside a  region of just a few arcsec (i.e., ${\rm r} = 2''$).
Although this value is slightly higher than the one derived for the $3'' \times 3''$ region of this galaxy (see Table~\ref{t_mass}), these measurements are consistent within the errors. The difference is due to the different approaches followed in the determination of the continuum of the spectrum in each region.

The ranges of gas masses found in the nucleus of the galaxies in our sample are comparable to the range of \Mbh\ usually found in the centre of galaxies.
From the stellar velocity dispersion $\sigma$ and the rotational velocity V profiles available in the literature, it is possible to estimate the \Mbh\ of the galaxies using the ${\rm M}_{\rm BH}-\sigma_e$ \citep[as in][]{Gultekin2009}.
Table~\ref{t_mbh} lists the values of $\sigma_{\rm e}=(\sigma^2+{\rm V}^2)^{1/2}$ derived for the galaxies in our sample and the corresponding \Mbh\ assuming the ${\rm M}_{\rm BH}-\sigma$ relation given by \citet{McConnell2011a} for spiral galaxies.
The estimated \Mbh\ are about an order of magnitude higher than the mass of the gas contained inside a region of ${\rm r}=0.2''$ for all the galaxies except \n4579 (see Fig.~\ref{f_enclosed}). As we mentioned above, the latter shows a particularly high content of gas.
The galaxy \n4501 harbours the largest BH mass from the all sample, which is consistent with the value $ {\rm M_{BH}} = 2.7\times10^7$\Ms\ obtained from the dynamical modelling of the stars following the method described by \cite{Nowak2007}, which is based on the Schwarzschild superposition code of \cite{Thomas2004}. The full analysis of the stellar dynamical modelling of \n4501 will be presented in Erwin et al. (in preparation).

\begin{table}
\centering
\caption{Estimated black hole masses for the galaxies in the sample.}
\label{t_mbh}
\begin{tabular}{lcc}
\hline
Galaxy & $(\sigma^2 + V^2)^{1/2}$  & \Mbh$^*$ \\
       & [\kms] & [$10^6$\Ms] \\
\hline
\n3351 & $100.9 \pm 9.1$ &  4.0 (2.6--6.0)\\
\n3627 & $122.4 \pm 7.3$ &  9.8 (7.4--12.8) \\
\n4501 & $157.8 \pm 36.9$ & 31.5 (92.9--82.6)   \\
\n4536 & $103.1 \pm 14.5$ & 4.5 (2.2--8.2) \\
\n4569 & $114.2 \pm 13.6$ & 7.2 (4.0--12.0)   \\
\n4579 & $140.2 \pm 16.0$&  18.3 (10.5--30.1) \\
\hline 
\end{tabular}
\\
$^{*}$ The numbers between parenthesis correspond to the lower and upper limits of the \Mbh\ given by the errors in sigma. 
\end{table}

But how large is the gas mass in comparison to the stellar mass, and how much can it affect the \Mbh\ dynamical estimates? To answer this question we have included in Fig.~\ref{f_enclosed} the de-projected spherical stellar masses as a function of the radius for the galaxies \n4501, \n4536, \n4569 and \n4579.
These distributions were obtained from the luminosity profile and the M/L ratio derived for each galaxy.
The luminosity density profiles were obtained by deprojecting surface-brightness profiles using a modified version of the algorithm of \cite{Magorrian1999}, taking into account the effects of PSF convolution \citep[see][]{Rusli2011}. The surface-brightness profiles (typically in near-IR bands such as $H$ and $K$) were derived from a combination of ellipse fits to isophotes and 2D modeling of images, using high-resolution HST images for the smallest scales and SDSS or Spitzer IRAC images for larger scales; in cases where only optical HST images were available and strong near-nuclear dust lanes were present, we also used $K$-band images derived from our SINFONI datacubes. Details for individual galaxies, along with the dynamical modeling, will be discussed in Erwin et al. (in preparation).
In the case of \n4501, the M/L ratio was obtained from stellar dynamical modelling. 
For the other galaxies (\n4536, \n4569 and \n4579, for which dynamical modeling is not yet complete), we derived the M/L ratio from the colours of the galaxies and the transformations given by \cite{Zibetti2009} and \cite{Bell2001}. These M/L are based on scaled Salpeter and Chabrier IMFs for \cite{Bell2001} and \cite{Zibetti2009}, respectively. Recently, evidence for a more dwarf-dominated IMF in hot stellar systems has grown \citep[e.g.][]{Thomas2011, Cappellari2012,Conroy2012, Dutton2012}. Hence, our M/L should be considered as lower limits for the actual stellar masses. From the analysis of the inner 5\arcsec\ of the galaxies in WFPC2 archive images, we derived average colours of $B-V = 1.0$ for \n4536, $V-I = 1.2$ for \n4569 and $V-I = 1.3$ for \n4579. These values are in very good agreement with the ones reported in the photometric compilation of \cite{Prugniel1998}  for the smallest ($\sim 6.6''$) aperture, and correspond to M/L$_{K}=0.6$, M/L$_{H}=1.2$ and M/L$_{K}=1.0$ for \n4536, \n4569 and \n4579, respectively.

It is clear from Fig.~\ref{f_enclosed} that the gas content of these galaxies is negligible compared to the stellar mass, which exceeds the former by more than an order of magnitude. In all the galaxies, the gas mass is no more than 0.1--3 per cent of the stellar mass, and even though the galaxies have considerable amounts of gas present in their nuclei, it is not enough to affect the dynamical \Mbh\ measurements based only on the observed stellar features of the galaxy.

\section{Summary and conclusions}\label{s_summary}

We have presented new AO-assisted $K$-band \S/VLT IFS observations of a sample of six nearby galaxies, \n3351, \n3627, \n4501, \n4536, \n4569 and \n4579. This is the first of two companion papers analysing the gas properties of these galaxies, and is part of a series reporting the results of our \S\ survey aimed at expanding the number of reliable BH masses in the nucleus of galaxies by means of dynamical modelling.

The \S\ spectra of the galaxies in the sample display several \H2\ emission lines and, in some cases, ionized hydrogen and helium emission-lines. This, together with the high-spatial resolution provided by the observations, allowed us to study in detail the morphology and physical properties of the emission-line gas located in the innermost regions of the galaxies, and the star-forming regions associated with it. Additionally, we have derived the gas mass content of the galaxies and compared it with their stellar masses in order to check if it could affect the BH masses derived from stellar dynamical modelling.
Our main results are summarized below. 
\begin{enumerate}
\item A rich diversity of morphologies is seen in the \H2\ emission-line gas distributions of galaxies, including bar- and ring-like structures, centrally concentrated emission and off-centre irregular distributions. When detected, the emission lines from ionized gas (i.e., \Brg\ and \HeI) showed also a variety of morphologies, not necessarily coincident with the \H2\ emission-line distribution.

\item \H2\ emission-line ratios point towards thermal processes as the main mechanism responsible for the excitation of the \H2\ emission-line gas, independent of the type of nuclear activity of the galaxies.
However, non-thermal processes are not completely ruled out, and seem to have a relative stronger contribution in the regions closer to AGNs. This scenario is supported by the measured \H2/\Brg\ ratios, which are higher in the nuclear regions of the galaxies.

\item The high-spatial resolution provided by the \S\ data allowed us to resolve the circumnuclear rings traced by ionized gas in the central regions of \n3351 and \n4536. The analysis of the \Brg\ equivalent width indicates that the star formation in the ring of \n3351 and \n4536 started 6.9 and 6.5~Myr ago, respectively. Additionally, the \HeI/\Brg\ emission-line ratio measured in \n4536 is approximately constant, suggesting that the star formation started at the same time along the entire ring.

\item We found that a rescaling of the \H2~2.12\micron\ emission-line luminosity by a factor $\beta \simeq 1200$ gives a good estimate (within a factor of 2) of the total (cold) molecular gas mass.

\item The galaxies in the sample contain large quantities of molecular gas, with total masses in the range $\sim 10^5 - 10^8$\Ms. Never the less, these masses are only a 3 per cent (at most) of the corresponding stellar masses. Hence, BH mass estimates based on the dynamical modelling of the stars should not be affected by the presence of gas.

\end{enumerate}

\section*{Acknowledgments}

We thank the referee for his/her constructive comments on our manuscript. We thank the Paranal Observatory Team for support during the observations. SPR acknowledges support from the DFG Cluster of Excellence `Origin and Structure of the Universe'. PE was supported by the Deutsche Forschungsgemeinschaft through the Priority Programme 1177 'Galaxy Evolution'.
This research has made use of the NASA/IPAC Infrared Science Archive (IRSA) and the NASA/IPAC Extragalactic Database (NED), which are operated by the Jet Propulsion Laboratory, California Institute of Technology, under contract with the National Aeronautics and Space Administration.

\appendix

\section{}\label{appendix}

In this section we specify the location of the spatial regions from where we extracted the 1D spectra analysed in Section~\ref{s_pconditions}. The 1D spectra were obtained by summing up the signal inside elliptical apertures or the entire FOV (e.g., $\sim 3''\times 3''$ for the HR data). Table~\ref{t_regions} lists the centre, sizes of the semi-major and -minor axis and P.A. of the elliptical apertures. The P.A. is defined as the angle between the positive x-axis of the datacube and the semi-major axis of the region (measured anticlockwise). The spectrum of the regions called `Ring' of \n4501 and \n4536 corresponds to the region between two ellipses. In these cases we report the minimum and maximum semi-major and -minor axis. For the region `5 -- Ring' of \n3351 we summed up the signal of the entire FOV outside the elliptical aperture specified in Table~\ref{t_regions}.

\begin{table*}
\caption{Position and shape of the elliptical apertures used to extract the 1D spectra analysed in Section~\ref{s_pconditions}.}
\label{t_regions}
\begin{tabular}{lcccc}
\hline
\multirow{2}*{Galaxy regions} & \multirow{2}*{Centre} & semi-major & semi-minor  & \multirow{2}*{P.A.}\\
 &        & axis  & axis  & \\
\hline
\n3351 &&&&\\
\hspace{0.2cm} 1 -- Nucleus 		& $(0'',0'')$ & 0.75\arcsec\ & 0.55\arcsec\ &  90\deg\ \\
\hspace{0.2cm} 2 -- W cloud 		& $(1.05'',-0.15'')$ & 0.3\arcsec\ & 0.3\arcsec\ & 0\deg\ \\
\hspace{0.2cm} 5 -- Ring 	         & $(0'',0'')$ & $1.15''- $ & $1.15''-$ & 0\deg\ \\
\n3627 &&&&\\
\hspace{0.2cm} 1 -- Nucleus 		& $(0'',0'')$ & 1.0\arcsec\ & 0.75\arcsec\ & 90\deg\ \\
\hspace{0.2cm} 3 -- N cloud            & $(-3'',5.6'')$ & 0.87\arcsec & 0.87\arcsec & 0\deg  \\
\hspace{0.2cm} 4 -- S cloud            & $(2.1'',-3.75'')$ & 0.87\arcsec& 0.87\arcsec & 0\deg \\
\n4501 &&&&\\
\hspace{0.2cm} 1 -- Nucleus 		& $(0.2'',0'')$ & 0.85\arcsec &0.6\arcsec & 45\deg \\
\hspace{0.2cm} 2 -- Ring		& $(0.2'',0'')$ &  $0.85''- 2.25''$ &  $0.6''- 1.9''$& 45\deg \\
\n4536 &&&&\\
\hspace{0.2cm} 1 -- Nucleus 		& $(0'',0'')$ & 1.05\arcsec & 0.7\arcsec& 135\deg \\
\hspace{0.2cm} 2 -- Ring 	 	& $(0.25'',-0.125'')$ &  $1.875''- 5.375''$  &  $0.5''- 1.25''$  & 90\deg \\
\n4579 &&&&\\
\hspace{0.2cm} 1 -- Nucleus 		& $(-0.15'',-0.1'')$ & 0.8\arcsec & 0.7\arcsec & 135\deg \\
\hspace{0.2cm} 2 -- NE cloud 		& $(0.5'',0.8'')$ & 0.75\arcsec & 0.4\arcsec& 150\deg\\
\hline
\end{tabular}
\end{table*}

\end{document}